\documentclass[manuscript]{acmart}

\usepackage{graphicx}%
\usepackage{multirow}%

\usepackage{amsmath,amssymb,amsfonts}
\usepackage{amsthm}%
\usepackage{mathrsfs}%
\usepackage[title]{appendix}%
\usepackage{xcolor}%
\usepackage{xcolor}%
\usepackage{colortbl}
\usepackage{textcomp}%
\usepackage{manyfoot}%
\usepackage{algorithm}%
\usepackage{algorithmicx}%
\usepackage{algpseudocode}%
\usepackage{listings}%
\usepackage{longtable}
\usepackage{pifont}

\usepackage{tabularx}
\usepackage{makecell}

\usepackage{tablefootnote}
\usepackage{textcomp}
\usepackage[frozencache,cachedir=.]{minted}
\usepackage{hyperref}
\usepackage{enumitem}           
\usepackage[T1]{fontenc} 
\usepackage{tikz} 
\usepackage{stfloats}
\usepackage[utf8]{inputenc}
\usepackage{soul}
\usepackage{colortbl}
\definecolor{mygray}{rgb}{0.71,0.71,0.71}
\usepackage{balance}
\usepackage{amsfonts}
\usepackage[caption=false]{subfig} 
\usepackage{array}

\usepackage{mdframed}

\newcolumntype{P}[1]{>{\centering\arraybackslash}p{#1}}

\newcommand{\ie}{\textit{i.e.},\ }
\newcommand{\eg}{\textit{e.g.},\ }
\newcommand{\etal}{\textit{et al.} }
\newcommand{\etc}{{\em etc.}}

\usepackage{booktabs}

\definecolor{francBlue}{RGB}{64,76,87}

\usepackage[most]{tcolorbox}
\usepackage[parfill]{parskip}

\tcbset{
  parskip/.style={
    before={\par\pagebreak[0]\parindent=0pt},
    after={\parfillskip=0pt\par},
  },
}
\newtcolorbox{resultbox}[1][]{%
    colback=black!3,
    colframe=black!3,
    notitle,
    sharp corners,
    borderline west={2pt}{0pt}{gray!80!black},
    enhanced,
    breakable,
    boxsep=0pt,
    left=4pt,right=2pt,top=2pt,bottom=2pt,
    }
\newcommand{\rques}[1]{  
\begin{tcolorbox}[enhanced jigsaw,colback=white,left=0pt,right=0pt,top=0pt,bottom=0pt]
\textbf{#1}
\end{tcolorbox}
}

\definecolor{codebg}{rgb}{0.99,0.99,0.99}
\definecolor{hiliteColor}{rgb}{1,0.92549019607,0.6}
\definecolor{tainted}{rgb}{0,1,1}

\newmintinline{python}{fontsize=\normalsize}

\newminted[PythonSourceCode]{python}{
    labelposition=topline,
    frame=single,
    numbersep=2pt,
    fontsize=\tiny,
    xleftmargin=5pt,
    framerule=0.5pt,
    linenos
}

\newminted[JavaSourceCode]{java}{
    labelposition=topline,
    frame=single,
    numbersep=2pt,
    fontsize=\tiny,
    xleftmargin=5pt,
    framerule=0.5pt,
    linenos
}

\definecolor{magnolia}{rgb}{0.97, 0.96, 1.0}
\definecolor{shadecolor}{rgb}{0.97, 0.96, 1.0}
\newcommand{\code}[1]{\texttt{\small{#1}\normalsize}}
\newmintinline[codePython]{python}{fontsize=\small}
\newmintinline[codeJava]{java}{fontsize=\small}
\newmintinline[snippetJava]{java}{fontsize=\small,bgcolor=magnolia}
\newmintinline[snippetPython]{python}{fontsize=\small,bgcolor=magnolia}
\newmintinline[snippetPerl]{perl}{fontsize=\small,bgcolor=magnolia}

\newcommand{\joanna}[1]{\textcolor{magenta}{}}
\newcommand{\latif}[1]{\textcolor{red}{#1}}

\setlist{nosep, topsep=0pt, partopsep=0pt, parsep=0pt, itemsep=0pt}

\setlist[itemize]{noitemsep, topsep=0pt, leftmargin=*}

\setcopyright{acmcopyright}
\copyrightyear{2026}
\acmYear{2026}
\setcopyright{acmlicensed}

\acmDOI{XXXXXXX.XXXXXXX}

\acmJournal{TOSEM}

\acmISBN{978-1-4503-XXXX-X/18/06}

\begin{document}

\title{An Empirical Study on Remote Code Execution in Machine Learning Model Hosting Ecosystems}

\author{Mohammed Latif Siddiq}
\email{msiddiq3@nd.edu}
\orcid{0000-0002-7984-3611}
\affiliation{%
  \institution{University of Notre Dame}
  \city{Notre Dame}
  \state{IN}
  \country{USA}
  \postcode{46556}
}

\author{Tanzim Hossain Romel}
\email{romel.rcs@gmail.com}
 \orcid{0009-0009-2432-8960}
\affiliation{%
  \institution{IQVIA Inc}
  \city{Durham}
  \state{NC}
  \country{USA}
  \postcode{27703}
}
\author{Natalie Sekerak}
\email{nsekerak@nd.edu}
\orcid{0009-0005-6626-1961}
\affiliation{%
  \institution{University of Notre Dame}
  \city{Notre Dame}
  \state{IN}
  \country{USA}
  \postcode{46556}
}

\author{Beatrice Casey}
\email{bcasey6@nd.edu }
\orcid{0009-0001-0097-2120}
\affiliation{%
  \institution{University of Notre Dame}
  \city{Notre Dame}
  \state{IN}
  \country{USA}
  \postcode{46556}
}
\author{Joanna C. S. Santos}
\email{joannacss@nd.edu}
\orcid{0000-0001-8743-2516}
\affiliation{%
  \institution{University of Notre Dame}
  \city{Notre Dame}
  \state{IN}
  \country{USA}
  \postcode{46556}
}


\begin{abstract}

Model-sharing platforms, such as Hugging Face, ModelScope, and OpenCSG have become central to modern machine learning development, enabling developers to share, load, and fine-tune pre-trained models with minimal effort. However, the flexibility of these ecosystems introduces a critical security concern: the execution of untrusted code during model loading (\ie via \texttt{trust\_remote\_code} or \texttt{trust\_repo}). In this work, we conduct the first large-scale empirical study of custom model loading practices across five major model-sharing platforms to assess their prevalence, associated risks, and developer perceptions. 
We first quantify the frequency with which models require custom code to function and identify those that execute arbitrary Python files during loading. We then apply three complementary static analysis tools: Bandit, CodeQL, and Semgrep, to detect security smells and potential vulnerabilities, categorizing our findings by CWE identifiers to provide a standardized risk taxonomy. We also use YARA to identify malicious patterns and payload signatures. In parallel, we systematically analyze the documentation, API design, and safety mechanisms of each platform to understand their mitigation strategies and enforcement levels. Finally, we conduct a qualitative analysis of over 600 developer discussions from GitHub, Hugging Face, and PyTorch Hub forums, as well as Stack Overflow, to capture community concerns and misconceptions regarding security and usability. Our findings reveal widespread reliance on unsafe defaults, uneven security enforcement across platforms, and persistent confusion among developers about the implications of executing remote code. 
We conclude with actionable recommendations for designing safer model-sharing infrastructures and striking a balance between usability and security in future AI ecosystems.
\end{abstract}





\maketitle
\section{Introduction}
Large Language Models (LLMs) have been increasingly used in day-to-day conversation and assisting tasks \cite{brown2020language,ouyang2022training,openai2023gpt4}. These models are based on 
different transformer architectures \cite{attention2017} and their advancements. These have enabled the creation of models with unprecedented scale, often comprising billions or even trillions of parameters \cite{openai2023gpt4}. Models are continuously reused, re-tuned, and evaluated for new tasks. \textbf{Model hubs} (or model registries) like Hugging Face play a critical role in this ecosystem by providing a centralized platform for hosting and sharing pre-trained models and datasets \cite{Jian24modelsarecode}. As of January 2026, Hugging Face alone hosts around 2.1 million models, fostering an open and collaborative environment for developers and researchers worldwide.

While model hubs and their supporting libraries (\eg \texttt{transformers} \cite{wolf-etal-2020-transformers} and \textsf{PyTorch} \cite{pytorch2025}), enable the seamless distribution of model weights, some models inherently require code execution to function correctly \cite{huggingface_transformers_custom_models}. Early neural network architectures relied on standardized, composable layers that could be fully described through configuration files. In contrast, current LLMs often introduce non-standard architectural components (\eg custom attention mechanisms, domain-specific preprocessing steps, and hardware-aware optimizations) that cannot be easily serialized without accompanying executable code~\cite{Jian24modelsarecode}. For example, when researchers develop a new transformer variant with a novel positional encoding scheme, they must distribute not only the trained weights but also the accompanying Python code that specifies how those weights are applied during inference. Without this code, downstream users would need to manually re-implement the architecture, making model sharing inefficient and, in many cases, impractical~\cite{hu2025llmsupplychain}.

Allowing code to run during model loading increases flexibility but also introduces security risks~\cite{Jian24modelsarecode}. One security issue arises from unsafe serialization formats such as Python’s \texttt{pickle}, which can execute arbitrary code during deserialization via the \texttt{\_\_reduce\_\_} method, turning model files into potential attack payloads~\cite{kellas2025pickleball,casey2025empirical}. Prior work has shown that malicious pickle-based models in the wild have been used to deploy reverse shells and steal credentials~\cite{Jian24modelsarecode,casey2024large}.
Another  distinct security problem comes from \textit{custom remote code execution}, enabled when users load models with flags such as \code{trust\_remote\_code=True}. This allows arbitrary Python modules provided by model authors to run locally, extending the attack surface beyond serialized data to unverified source code.

Consequently, loading models from public hubs creates implicit trust relationships among users, model authors, and platforms (a trust that is often misplaced~\cite{jiang2022artifacts}). While previous work has examined deserialization attacks~\cite{hfpicklescan2024}, \textbf{\textit{the prevalence and risks of custom remote code execution during model loading remain largely unexplored}}.
Therefore, this paper closes this gap through a large-scale empirical study across five major model-sharing platforms (Hugging Face~\cite{huggingface2025}, OpenCSG~\cite{opencsg2025}, ModelScope~\cite{modelscope2025}, OpenMMLab~\cite{openmmlab2025}, and PyTorch Hub~\cite{pytorch2025}). We first quantify how often models rely on custom loading code and identify those that execute arbitrary Python files. We then apply multiple static analysis tools (Bandit, CodeQL, and Semgrep) to detect potential vulnerabilities and categorize them by CWE identifiers. In parallel, we analyze each platform’s documentation, APIs, and security controls to assess mitigation practices and qualitatively examine over 600 developer discussions from GitHub, Hugging Face, PyTorch Hub, and Stack Overflow to capture community perceptions and misconceptions about security and usability.

\subsection{Manuscript Contributions}
This work makes the following contributions:
The contributions of this work are:
\begin{itemize}[leftmargin=*]
    \item The first cross-platform, large-scale measurement study of untrusted model code execution across five major model-sharing platforms: Hugging Face \cite{huggingface2025}, ModelScope \cite{modelscope2025}, OpenCSG \cite{opencsg2025}, OpenMMLab \cite{openmmlab2025}, and PyTorch Hub \cite{pytorch2025}.
    \item We systematically detect and categorize security weaknesses using three static analyzers in around 45,000 repositories containing custom code. Moreover, we incorporate signature-based malicious pattern detection using YARA \cite{yara} to identify potential payloads.
    \item We analyze platform-level defenses, including warning systems, static and dynamic scanning, and trust flag mechanisms.
    \item We create a taxonomy about developers' perception about remote code execution during model loading after examining around 600 developer discussions from forums, GitHub issues, pull requests, and Q\&A sites.
\end{itemize}

\subsection{Manuscript Organization}
The remainder of this manuscript is organized as follows. Section~\ref{sec:background} introduces the necessary background on model loading with executable code, code smells, and the threat model used in our work. Section \ref{subsec:ThreatModel} provides the threat model used in our study. Section~\ref{sec:rqs} provides the research questions answered in this study. Section~\ref{sec:methodology} describes our dataset collection and quantitative and qualitative analyses. Section~\ref{sec:results} presents the results addressing each research question. Section~\ref{sec:discussion} discusses the implications of our findings, limitations, and directions for future work. Section \ref{sec:related} provides related literature and a comparison with our work. Finally, Section~\ref{sec:conclusion} concludes the paper.

\section{Background}
\label{sec:background}

This section establishes the fundamental concepts necessary to understand the work.

\subsection{Model Loading with Executable Code}
\label{subsec:code_execution}

Unlike traditional data files, modern ML models may require code execution for technical reasons. Early neural networks consisted of standardized layers that could be described solely by configuration, but contemporary architectures implement novel mechanisms (\eg custom attention patterns, domain-specific preprocessing, \etc) that cannot be expressed without executable code~\cite{Jian24modelsarecode}. Thus, when developers implement a new transformer variant with unique positional encoding, they must ship \textit{both} the trained weights and the Python code defining how those weights interact. The alternative would require every user to manually reconstruct the architecture, making large-scale model sharing impractical~\cite{hu2025llmsupplychain}.



% ...


This technical need manifests through platform APIs that are simple to use. When developers call \code{from\_pretrained} or \code{pipeline} methods, the \code{transformers} library downloads multiple files, including Python modules that execute with full system privileges when remote code trust flags are enabled. Whenever the repository contains specific entry-point files, such as \texttt{modeling\_*.py}, \texttt{tokenizer.py}, or \texttt{hubconf.py}, they are automatically imported and executed as part of the model initialization process. While these files often contain legitimate code that defines how the model operates,  an attacker can embed malicious payloads in them that would execute with the same privileges as any other local Python process. This means that enabling \code{trust\_remote\_code} (for \code{transformers}) or \code{trust\_repo} (for PyTorch Hub) effectively grants remote repositories the ability to run arbitrary Python code on the host machine. Platforms like Hugging Face host over 2.1 million models, with thousands added daily, making manual review impractical and automated scanning insufficient~\cite{laufer2025anatomy}. 

\begin{figure}[!htbp]
    \centering
    \includegraphics[width=\linewidth]{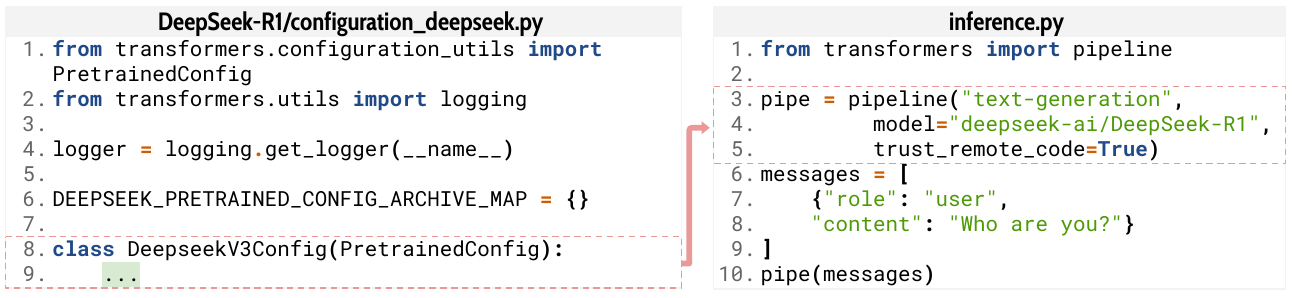}
    \caption{Example of a custom configuration script (left) and loading a custom model (right).}
    \label{fig:custom_model}
\end{figure}

To illustrate, Figure~\ref{fig:custom_model} shows how model loading can trigger code execution. On the left, a developer defines a custom configuration class (\code{DeepseekV3Config}) that extends the \code{transformers} library’s base configuration. This file, stored in the model repository, contains Python code that can be executed when the model is loaded. On the right, a user loads this same model from Hugging Face using the \code{pipeline} API and sets \code{trust\_remote\_code=True}. This flag tells the library to trust and run any Python code provided by the remote repository, effectively downloading and executing unverified scripts from the internet.

This customization mechanism introduces a serious security risk: an attacker can upload a model repository containing \textbf{malicious code} to deliver payloads like reverse shells, keyloggers, or data exfiltration scripts~\cite{Jian24modelsarecode}. Since the customization code executes automatically during model loading, users who enable \code{trust\_remote\_code} may unknowingly grant full system access to untrusted code. This behavior makes large-scale model sharing both powerful and potentially dangerous, especially when combined with the high volume of new models uploaded daily.

\subsection{Code Smells \& Security Smells}


\textbf{\textit{Code smells}} are indicators of poor design or implementation choices that may not immediately cause failures but often lead to maintainability issues and increased defect risk~\cite{fowler1999,pereira22}. They typically reflect violations of good design principles, making software harder to evolve and more error-prone. 
For example, in Listing~\ref{lst:smell} (left), the ValueError exception is caught but not handled—an anti-pattern that reduces error transparency and can lead to subtle failures~\cite{gupta2018software}.

\begin{listing}[!ht]
\caption{Examples of a code smell (left) and a security smell (right).}
\label{lst:smell}
\noindent\begin{minipage}{.495\linewidth}
\begin{PythonSourceCode*}{highlightlines=4-5,label=\tiny{\textsf{Code smell example}}}
try:
    num = input('Enter number:')
    num = int(num)
except ValueError:
    pass
\end{PythonSourceCode*}
\end{minipage}\hfill
\begin{minipage}{.495\linewidth}
\begin{PythonSourceCode*}{highlightlines=3-3,label=\tiny{\textsf{Security smell example}}}
import hashlib
def validate(c, h):
    hash_md5 = hashlib.md5(c)
    hash = hash_md5.hexdigest()
    return hash == h
\end{PythonSourceCode*}
\end{minipage}
\end{listing}

A specific subset of code smells, called \textbf{\textit{security smells}}, is associated with patterns that may introduce or signal the presence of vulnerabilities~\cite{rahman_seven_2019,rahman2019share,ghafari2017security}. These patterns do not always constitute exploitable vulnerabilities but highlight code areas where security controls are weak or outdated.  For example, in Listing~\ref{lst:smell} (right), the use of the insecure \texttt{md5} algorithm is associated with \textit{CWE-327: Use of a Broken or Risky Cryptographic Algorithm}~\cite{siddiq2022empirical}. Such patterns increase the likelihood of exploitation if left unaddressed.

\section{Threat Model}\label{subsec:ThreatModel}

\begin{figure}[!htbp]
    \centering
    \includegraphics[width=\linewidth]{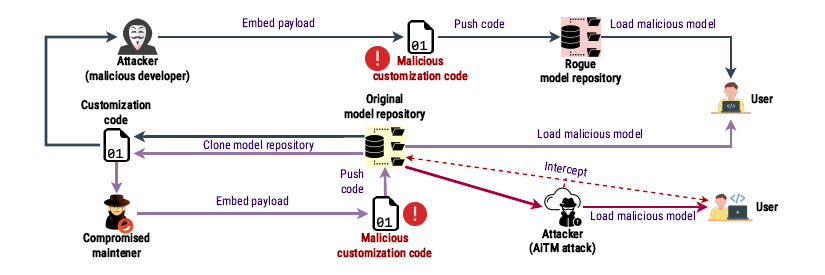}
    \caption{Threat model overview}
    \label{fig:threat}
\end{figure}

Custom model loading introduces complex trust boundaries among three main stakeholders: \textit{model creators}, \textit{platform maintainers}, and \textit{model consumers}. Figure \ref{fig:threat} provides an overview of our threat model, which examines how these boundaries can be exploited when platforms allow arbitrary code execution during model loading.

\subsection{Adversary Assumptions}
We assume that attackers have accounts on model platforms and can upload models or modify existing repositories. Adversaries may also leverage community features, such as discussions or pull requests, to disseminate or promote malicious content. We consider two  adversary types:
(1)~a \textit{malicious developer} who intentionally uploads a model with harmful customization code, and
(2)~a \textit{compromised maintainer} whose credentials or tokens were hijacked to distribute poisoned models under the guise of a reputable author.
Importantly, not all threats arise from deliberate attacks. Model contributors who are unaware of secure publishing practices may unintentionally introduce insecure code, effectively expanding the threat surface without adversarial intent.

\subsection{Threat Scenarios}
Three threat scenarios exist (\textbf{S1}–\textbf{S3}):
\begin{itemize}[leftmargin=15pt]
\item[\textbf{S1}] \textbf{Malicious Fork.}
An attacker downloads an existing benign model $M$, changes the custom code with a harmful payload (\eg~reverse shell), and uploads a modified model $M'$ to their own (rogue) repository while presenting it as an enhanced or compatible version of the original. Model consumers would need to set \texttt{trust\_remote\_code=True} to be able to use the model, which would lead to the execution of the malicious payload.

\item[\textbf{S2}] \textbf{Compromised Trusted Account.}  
A trusted maintainer account is hijacked, allowing the attacker to upload a modified model $M'$ with malicious accompanying custom code directly to a legitimate, widely used repository. Consumers may trust the model due to its reputation, verified badges, or high download counts. Similar to Scenario~1, attackers can embed arbitrary Python logic in initialization files or leverage dependency manipulation. Social engineering and trust hijacking further increase the likelihood of exploitation, as users are more inclined to enable \texttt{trust\_remote\_code=True} for ``trusted'' sources.

\item[\textbf{S3}] \textbf{Attacker-in-the-middle (AiTM).}  
An attacker intercepts or tampers with the model distribution channel and modifies the model's custom code during transfer or dependency resolution (\eg~via compromised mirrors, registries, or proxy layers). This can also occur indirectly through poisoning or replacing cached artifacts stored by hosting platforms (\eg~Hugging Face cache directories), allowing the attacker’s modified version to be loaded even if the original upstream repository remains clean. As a result, when users enable \texttt{trust\_remote\_code=True}, they may execute the injected payload from the cached or intercepted model, effectively transforming a previously benign artifact into a malicious one.
Attackers may exploit dependencies and supply-chain manipulation, injecting malicious payloads at download or cache resolution time. Since the model is cached locally, future loads may execute the attacker’s payload even without further network interaction.


\end{itemize}

Not all risks stem from malicious actors. Model creators with limited security awareness may unintentionally include unsafe initialization code, hard-coded credentials, or insecure dependency calls. Although unintentional, such models can still be weaponized post-deployment, expanding the platform’s overall attack surface without deliberate adversarial behavior. Weak sandboxing, overly permissive dependencies, and a lack of static or runtime checks allow insecure code to run automatically during model loading, making these models soft targets for downstream exploitation.
\subsection{Trust Relationships.}
Model consumers implicitly trust platform interfaces and configuration defaults (\eg \texttt{trust\_remote\_code=True} or \texttt{trust\_repo=True}) to safely retrieve and execute model code. This trust is often amplified by perceived platform reputation or download counts, which may lead users to overlook warning prompts or disable security mechanisms for convenience. Platform maintainers, in turn, trust model contributors to follow safe publishing practices, while contributors depend on the platform to enforce isolation and verification mechanisms. The intersection of these assumptions creates a vulnerable trust boundary.



\section{Research Questions (RQs)}\label{sec:rqs}
As shown in Figure~\ref{fig:method}, we answer four RQs that explore the \textit{prevalence}, \textit{risks}, and \textit{developers' perceptions} of custom model loading across model hubs:      

\rques{\textbf{RQ1}: To what extent is custom model loading required?}  
Model-sharing platforms, such as Hugging Face and ModelScope, enable developers to provide custom code for loading or configuring models. While these capabilities increase flexibility and support novel architectures, they also introduce security risks within the ecosystem. In this RQ, we investigate how many models hosted on these platforms include custom code that is required to be executed upon model loading.

\rques{\textbf{RQ2}: Do remote code implementations for custom models contain vulnerabilities, security smells, or malicious payloads?}  
Allowing arbitrary code execution raises concerns about the introduction of vulnerabilities and insecure practices. Previous work has shown that code provided by the community may contain security smells, such as unsafe deserialization~\cite{Jian24modelsarecode,casey2024large,casey2025empirical}. In this RQ, we perform a systematic analysis of the models' customization code to determine the prevalence of vulnerabilities and potential exploit vectors. 

\rques{\textbf{RQ3}: What do the platforms offer for developers to mitigate the execution during model loading?}  
Platform hubs play a crucial role in enforcing safe defaults, providing comprehensive documentation, and implementing technical safeguards (\eg sandboxing, warning banners, or permission systems). In this RQ, we investigate the existing security mechanisms provided by various platforms. We examine whether platforms provide static or dynamic checks, whether they expose developers and end-users to explicit warnings when running custom code, and how policies such as \texttt{trust\_remote\_code} or \texttt{trust\_repo}  flags or isolated execution environments are enforced in practice.

\rques{\textbf{RQ4}: What are the developers’ concerns around code execution during model loading?}  
Beyond technical vulnerabilities, it is crucial to understand the perspective of developers who contribute to and use these models. Their concerns may range from usability (\eg friction in using security mechanisms) to trustworthiness (\eg fear of executing malicious code) and maintainability (\eg lack of long-term platform support for their contributions). In this RQ, we collect and analyze developer discussions from Hugging Face and PyTorch Hub forums, GitHub discussions, pull requests, issues, and StackOverflow Q\&A platforms to understand practitioners' concerns, misconceptions, and expectations.

\section{Methodology}\label{sec:methodology}
\begin{figure}[!htbp]
    \centering
    \includegraphics[width=\linewidth]{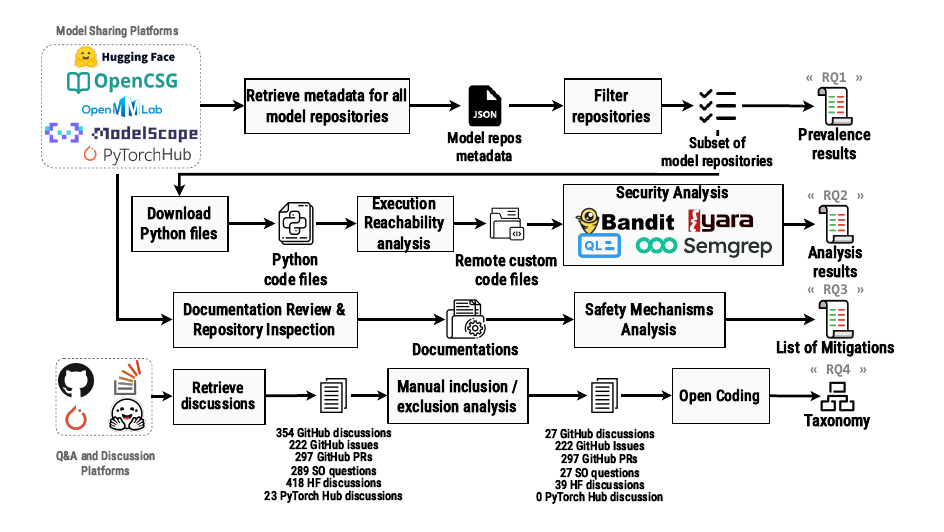}
    \caption{Methodology Overview}
    \label{fig:method}
\end{figure}

In this section, we provide details about the platform selection, data collection, and methodology to answer the research questions. Figure~\ref{fig:method} provides an overview of our methodology.

\subsection{Model Sharing Platform Selection}
We first examined a recent list of popular model-sharing platforms compiled by Jian \etal~\cite{Jian24modelsarecode}, which identified the top 15 model hubs. We then applied the following inclusion criteria to determine whether a given model hub (henceforth referred to as a ``platform'') qualified for our study. A platform was included only if it satisfied \textit{all} of the following conditions:
\textsf{(1)} it is publicly accessible; 
\textsf{(2)} it provides practical means to retrieve information from all hosted model repositories (\eg through APIs or web crawling); 
\textsf{(3)} the hosted models can be programmatically fetched and instantiated through standard model-loading APIs provided by \texttt{transformers}, \texttt{torch.hub}, or their respective wrappers, without requiring any reimplementation; and
\textsf{(4)} it supports libraries or mechanisms that permit the execution of custom code during model loading.
After this manual analysis, we identified \textbf{5} platforms for inclusion in our study: Hugging Face~\cite{huggingface2025}, OpenCSG~\cite{opencsg2025}, ModelScope~\cite{modelscope2025}, OpenMMLab~\cite{openmmlab2025}, and PyTorch Hub~\cite{pytorch2025}. 

\subsection{Model Repositories Retrieval and Selection}
After selecting the platforms, we collected metadata for all model repositories published within these platforms. Specifically, we captured the repository's access URLs, tags, and file list. 
Next, we applied the following inclusion criteria to identify model repositories that require custom code execution during model loading. A repository was included if it satisfied any of the following criteria:
\textbf{(i)} it was tagged with the \texttt{custom\_code} tag; or \textbf{(ii)} it contained one of the files \texttt{tokenizer.py}, \texttt{\_\_init\_\_.py}, or \texttt{hubconf.py}; or \textbf{(iii)} it included a Python file whose name began with \texttt{modeling\_}, \texttt{tokenization\_}, or \texttt{configuration\_}.
These conditions are based on the documentation of the \texttt{Pytorch Hub}~\cite{pytorch_hub_docs} and the \texttt{transformers} library~\cite{huggingface_transformers_custom_models}. The OpenMMLab platform, on the other hand, has its own library for loading custom models. Therefore, we included all the repositories listed on this platform in our analysis. 

\subsection{Execution-Relevant Reachability Analysis}
\label{subsec:jarvis}
Model repositories frequently include auxiliary scripts, legacy files, or utilities that are never invoked during model initialization. While such code may contain vulnerabilities, including it in our analysis would substantially overestimate practical security risk in this context. Therefore, we first retrieve \textit{all} Python scripts in them, and then we examine these scripts to identify only those that are \emph{execution relevant}, \ie Python code paths that may execute during model loading when \texttt{trust\_remote\_code=True}. 

To do so, we first identify model-loading entry points using the files described in the previous section for detecting custom code (\eg~\texttt{tokenizer.py}, \texttt{\_\_init\_\_.py}, and \texttt{hubconf.py}). These modules are treated as \textit{analysis roots}. Starting from these entry points, we use \textsf{JARVIS}~\cite{huang2024scalablepreciseapplicationcenteredgraph} to construct an interprocedural, flow- and context-sensitive call graph over the repository. \textsf{JARVIS} builds on \textsf{PyCG}~\cite{pycg} and enhances it with application-centered, context-sensitive analysis, enabling more precise resolution of Python-specific constructs such as class hierarchies, method dispatch, and imports. Thus, we compute the transitive closure of reachable functions and modules from the entry points and discard all code outside this reachable set. As a result, from an initial corpus of 128,487 Python files, our reachability analysis identifies 67,009 files as execution-relevant and includes only these files in subsequent security analyses, more specifically in RQ2.

\subsection{Security Smell, Vulnerability, and Malicious Payload Analysis}
After collecting the model repositories, we used three static analyzers to identify security smells and potential vulnerabilities: Bandit \cite{bandit2025}, CodeQL \cite{codeql2025}, and Semgrep \cite{semgrep2025}. To further identify malicious patterns and payload signatures, we employed YARA~\cite{yara}. These tools provide complementary coverage, combining lightweight static analysis with signature-based detection of malicious code.

\begin{itemize}

\item{Bandit (v1.8.6)} 
It is a security linter that statically inspects the abstract syntax tree (AST) of Python code to detect common vulnerabilities such as the use of unsafe functions (\texttt{eval}, \texttt{exec}, \texttt{pickle.load}), weak cryptographic algorithms, hardcoded credentials, and insecure temporary file creation. It also maps findings to \textit{Common Weakness Enumeration (CWE)}, which is a list of common types of software vulnerabilities \cite{cwe2025}. We executed Bandit recursively on all Python files extracted from each model repository, generating structured {JSON} outputs for aggregation and comparison. 

\item{CodeQL (v2.15.0)} 
It performs static analysis by compiling source code into a relational database of program elements (\eg functions, variables, control flow, and data flow) and executing declarative \texttt{QL} queries to detect security flaws. It enables inter-procedural and data-flow analysis for complex vulnerabilities such as injection, path traversal, and insecure deserialization. We executed CodeQL with the official query pack provided by GitHub Security Lab \cite{githubsecuritylab_codeql}.

\item{Semgrep (v1.139.0)} 
It is a lightweight, multi-language static analyzer that uses rule-based pattern matching to detect both general and domain-specific security issues. Unlike CodeQL, which requires query compilation, Semgrep matches syntactic and semantic patterns directly in the codebase, making it efficient for large-scale scanning. We used its built-in rulesets to identify security misconfigurations, unsafe API usage, and insecure imports across various model repositories. Findings were grouped by CWE and severity level to facilitate cross-tool comparison.

\item{YARA (v4.5.2)} 
It is a rule-based pattern-matching engine widely used in malware detection and digital forensics \cite{naik2020evaluating, naik2021embedded}. Unlike the previous static analyzers, which focus on identifying insecure coding patterns or API misuse, YARA detects known malicious behaviors through signature-based matching of strings, byte sequences, and regular expressions. This approach allows it to uncover embedded payloads that may not manifest as conventional security smells—for example, reverse shells, obfuscated network beacons, or credential-stealing scripts. The official YARA GitHub repository~\cite{virustotal_yara} provides a reference to a curated list of well-known rule sources~\cite{inquest_awesome_yara}. From these 70 publicly available sources, we successfully compiled 7,657 rules from 25 sources to scan the collected repositories (they failed to compile rules due to them being outdated with respect to the version we have used in our work).  
\end{itemize}

Beyond these static analyzers, we additionally evaluated security tools such as ClamAV~\cite{clamav}, PickleScan~\cite{picklescan}, ModelScan~\cite{modelscan}, and ModelAudit~\cite{modelaudit}. These tools are widely adopted by Hugging Face to assess the safety of uploaded models and datasets. However, they primarily operate on serialized model artifacts, focusing on detecting malicious payloads embedded within model files (\eg~pickled objects), rather than analyzing configuration files used to customize model behavior or the broader repository-level codebase. We executed these tools on our collected repositories but did not observe any findings within execution-relevant functions or modules reachable from the model-loading entry points identified in Section~\ref{subsec:jarvis}. Consequently, to ensure methodological consistency and comparability across platforms, we restrict our reported results to the previously described static analyzers (\eg~Bandit, CodeQL, \etc).


\subsection{Platform Mitigation}
\label{subsec:rq3_method}

To answer RQ3, we conducted a structured, multi-source analysis of platform-level mechanisms intended to mitigate unsafe model loading when custom code execution is permitted. Our analysis focused on both \emph{declared security policies} and \emph{practical enforcement mechanisms} implemented by each platform: Hugging Face, ModelScope, PyTorch Hub, OpenMMLab, and OpenCSG. Our scope was restricted to mechanisms that operate \emph{at the platform level} rather than user-controlled mitigations (\eg~local sandboxing or downstream deployment hardening). 
For each platform, we systematically reviewed official documentation, API references, security advisories, and public policy statements to identify declared trust models, upload constraints, and warnings related to executing remote code (\eg~\texttt{trust\_remote\_code} or \texttt{trust\_repo}). This step established the \emph{intended} security posture of each platform.

When source code or infrastructure components were publicly available, we examined open-source repositories and client-side libraries to verify whether documented safeguards were programmatically enforced. This included identifying:
(i) gating mechanisms requiring explicit user opt-in,
(ii) automated scanning pipelines (static or dynamic),
(iii) interactive confirmation prompts,
(iv) file-type restrictions (\eg~SafeTensors),
and (v) review-based admission controls (\eg~mandatory pull requests).
We distinguish between \emph{policy-level guarantees} and \emph{best-effort safeguards} that rely on user awareness or voluntary compliance.

To enable consistent comparison across platforms, we categorized mitigation mechanisms along four orthogonal dimensions:
\emph{(1) trust model} (open, community, or verify-first),
\emph{(2) automated safeguards} (\eg~malware scanning, pickle/RCE detection),
\emph{(3) user-facing protections} (warnings, prompts, badges, documentation), and
\emph{(4) execution isolation} (sandboxing or runtime containment).
Each platform was evaluated against all dimensions, even when a category was absent.

All claims were cross-validated against multiple independent sources (documentation, code repositories, and issue trackers) and reflect platform behavior as of October~2025. When enforcement could not be directly verified, we conservatively report only documented or observable behaviors and explicitly avoid inferring guarantees beyond what platforms publicly commit to.

\subsection{Developers’ Concerns}
To answer our RQ4, we focused on the Hugging Face and PyTorch Hub forums, GitHub discussions, pull requests, issues, and the Stack Overflow Q\&A platform. We searched these platforms using the keywords \texttt{trust\_remote\_code} and \texttt{trust\_repo}. We identified a total of \textbf{418} Hugging Face forum posts, \textbf{354} GitHub discussion posts, and \textbf{289} Stack Overflow posts. For GitHub issues, there were more than 13,000 retrieved issues, and for GitHub pull requests, more than 4,000 PRs, using the aforementioned query. Thus, we kept only issues and PRs that included any of the search query keywords in their titles. Then, two authors manually filter them in parallel to include them in our study based on their relevance. The Cohen's kappa score is 0.50, indicating a “moderate” level of agreement between the two authors \cite{cohen1960kappa}. A senior author with over 10 years of experience resolved the discrepancies. 

After manually filtering for relevance to our study, we retained 27 GitHub discussions, 222 GitHub issues, 297 GitHub pull requests, 27 Stack Overflow posts, and 39 Hugging Face discussions. An entry was deemed as relevant if it explicitly discussed the functionality, security implications, integration issues, maintenance actions, or community understanding related to \texttt{trust\_remote\_code} or \texttt{trust\_repo}, rather than merely mentioning the keyword in a code snippet as shown in the example in Listing \ref{fig:custom_model}. It is important to note that although we collected 23 discussions from PyTorch Hub's forums, none of them were relevant to our study.    

We then applied an \emph{open coding} approach to the selected posts~\cite{strauss1990basics}, carefully reading, analyzing, and annotating each post with conceptual labels (\emph{codes}) reflecting developers’ expressed sentiments, challenges, and misunderstandings. The coding was conducted collaboratively by the authors, whose software development experience ranged from 4 to 12 years. To ensure consistency, disagreements were discussed in weekly calibration meetings. If there was a discrepancy, the senior author mitigated it. We iteratively reviewed and refined the codes through regular calibration meetings until conceptual saturation was achieved.
\section{Results}
\label{sec:results}
\subsection{RQ1: Prevalence of Custom Models}
Table~\ref{tbl:rq1_results} provides a summary of the collected model repositories containing custom code. For Hugging Face, OpenCSG, and ModelScope, approximately 1\% to 4\% of the models include custom code. In contrast, for OpenMMLab and PyTorch Hub, all available repositories rely on custom code. 

\begin{table}[!htbp]
\centering
\scriptsize
\begingroup
\setlength{\aboverulesep}{0pt}
\setlength{\belowrulesep}{0pt}
\setlength{\tabcolsep}{4pt} 
\caption{Result of collected repositories with custom codes.}
\label{tbl:rq1_results}
\begin{tabular}{@{}lccc@{}}
\toprule
\textbf{Platform} & \textbf{\# Repos} & 
\begin{tabular}[c]{@{}c@{}}\textbf{\# Repos with} \\ \textbf{custom code}\end{tabular} & 
\textbf{(\%)} \\ 
\midrule
Hugging Face & 2,107,935 & 36,697 & 1.74\% \\ 
OpenCSG     & 192,556   & 6,165  & 3.20\% \\ 
ModelScope  & 68,736    & 3,193  & 4.65\% \\ 
OpenMMLab   & 16        & 16     & 100.00\% \\ 
PyTorch Hub & 26        & 26     & 100.00\% \\ 
\bottomrule
\end{tabular}
\endgroup
\end{table}

\begin{figure}[!htbp]
    \centering
    \includegraphics[width=\linewidth]{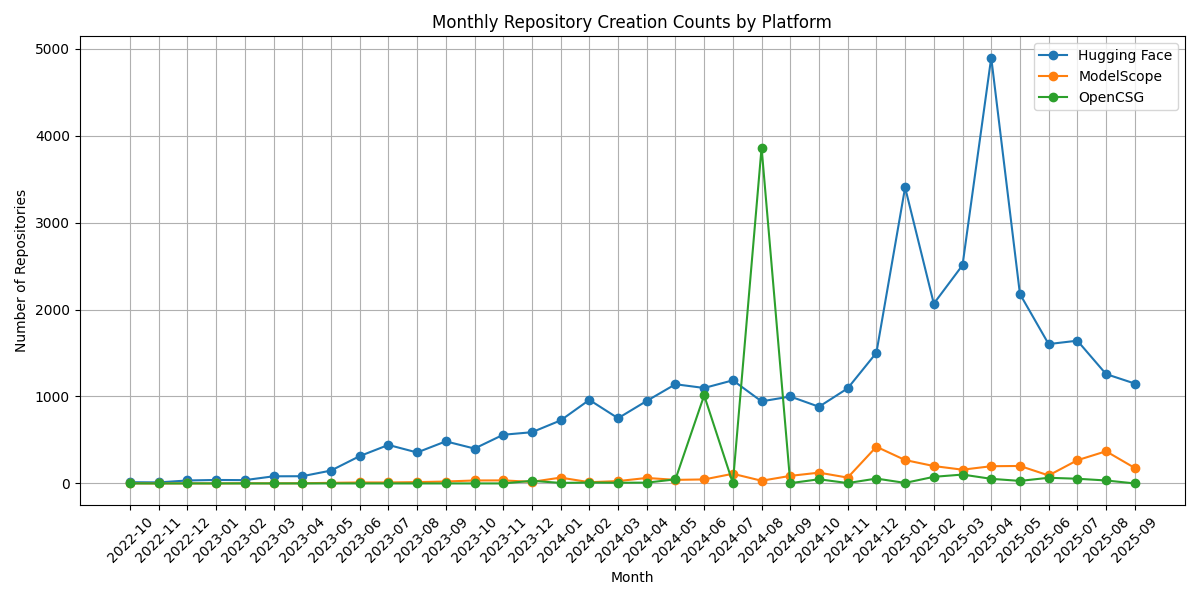}
    \caption{Monthly repository creations for custom models.}
    \label{fig:rq1_monthly}
\end{figure}

Figure \ref{fig:rq1_monthly} presents the monthly number of repositories created for hosting custom model-loading code across three major model hubs: Hugging Face, ModelScope, and OpenCSG, over a three-year period from October 2022 to September 2025. OpenMM and Pytorch Hub did not have any models created during this time period.

Across all platforms, we observe a general increase in the adoption of custom code over time; however, the growth trajectories differ substantially. \textbf{Hugging Face} exhibits sustained and large-scale growth throughout the entire observation window. Monthly repository creation steadily increases from double-digit counts in late 2022 to several thousand per month in 2025, peaking in April~2025 at 4{,}898 newly created repositories. This trend suggests that custom architectures, tokenizers, and preprocessing logic embedded directly within model repositories have become a normalized practice within the Hugging Face ecosystem.  The longitudinal data reveal a direct correlation between the release of "frontier" open-source models and the surge in repository creation on Hugging Face. The massive peak in \textbf{April 2025 (4,898 repositories)} and the sustained high volume in early 2025 can be attributed to several landmark releases. The January surge (3,405 repositories) aligns with the release of \textbf{DeepSeek-R1} \cite{deepseek_r1}. As one of the first open-source models to achieve parity with proprietary reasoning models (like OpenAI's o1 \cite{openai2024o1system}), it triggered a massive community effort to host "distilled" variants and specialized reasoning pipelines. The all-time peak in April coincides with a "dual-release" month. Meta launched the \textbf{Llama 4} family (including the \textit{Scout} and \textit{Maverick} models) on April 5, 2025 \cite{llama4_release}, while Alibaba Cloud unveiled \textbf{Qwen-3} on April 29, 2025 \cite{qwen3_release}. These releases shifted the ecosystem toward Mixture-of-Experts (MoE) architectures, necessitating thousands of new repositories for GGUF/EXL2 quantizations and vision-adapter code. A secondary surge in August 2025 (1,258 repositories) follows OpenAI's strategic shift to release \textbf{GPT-OSS}, their first open-weight model family (gpt-oss-120b and gpt-oss-20b) since GPT-2 \cite{gpt_oss_release}. This release under the Apache 2.0 license led to a spike in on-premises deployment configurations and agentic tool-calling repositories.


\textbf{ModelScope} shows moderate but consistent growth. Activity remains sparse during 2022 and early 2023, followed by a gradual increase throughout 2024 and a more pronounced expansion in late 2024 and 2025. Although its absolute volume remains substantially lower than Hugging Face's, the upward trend indicates an increasing reliance on custom execution pipelines within the platform. 

\textbf{OpenCSG} displays a highly volatile pattern characterized by extreme spikes followed by sharp declines. The platform remained relatively quiet until June 2024, when it surged to 1,013 repositories, then reached an all-time high of 3,863 in August 2024. This spike is not attributed to organic growth but rather to a \textit{platform-driven bulk event}. Specifically, the release of version 0.7.0 introduced \textbf{Multi-Source Mirroring} capabilities \cite{opencsg_v070}, allowing for the programmatic migration of thousands of external repositories. Furthermore, the August peak aligns with the platform's integration of \textbf{Model Context Protocol (MCP)} servers to support AgenticOps \cite{opencsg_agentic}.  The dramatic drops in activity observed in July and September 2024 (to 4 and 3 repositories, respectively) correspond to major infrastructure migrations. During September 2024, the platform migrated its portal to a Go-based architecture and implemented \textbf{Gitaly} for storage \cite{opencsg_v090}. Such structural overhauls typically involve a "write-lock" or a freeze on new repository creation to ensure data integrity, explaining the temporary cessation of new repository activity.

\begin{figure}[ht!]
    \centering
    \begin{minipage}[t]{0.19\textwidth}
        \includegraphics[width=\linewidth,keepaspectratio]{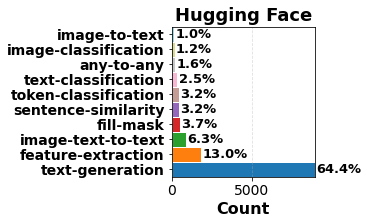}
    \end{minipage}\hfill
    \begin{minipage}[t]{0.19\textwidth}
        \includegraphics[width=\linewidth,keepaspectratio]{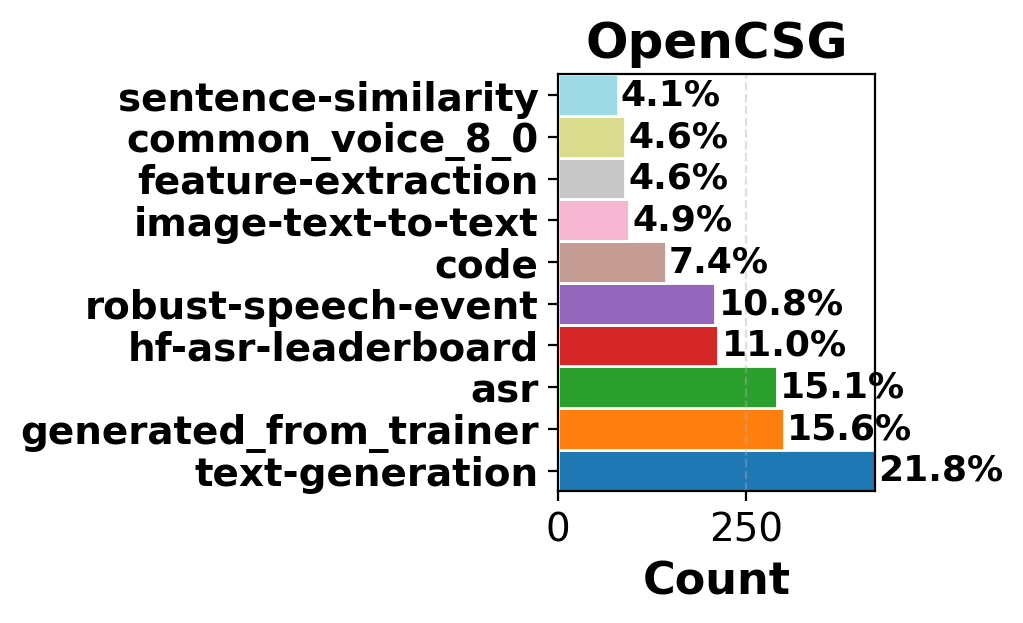}
    \end{minipage}\hfill
    \begin{minipage}[t]{0.19\textwidth}
        \includegraphics[width=\linewidth,keepaspectratio]{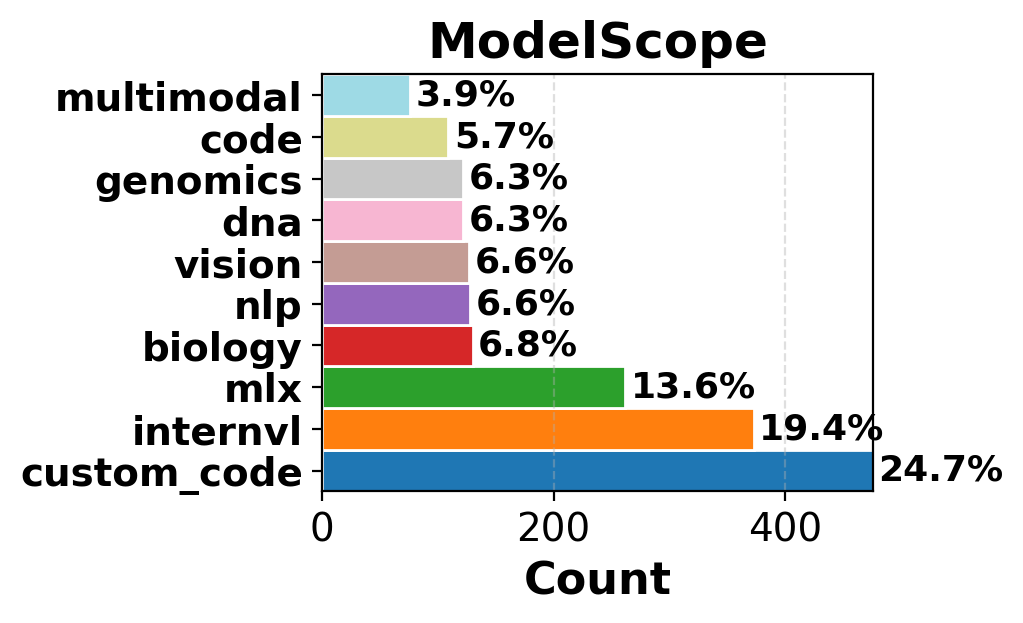}
    \end{minipage}\hfill
    \begin{minipage}[t]{0.19\textwidth}
        \includegraphics[width=\linewidth,keepaspectratio]{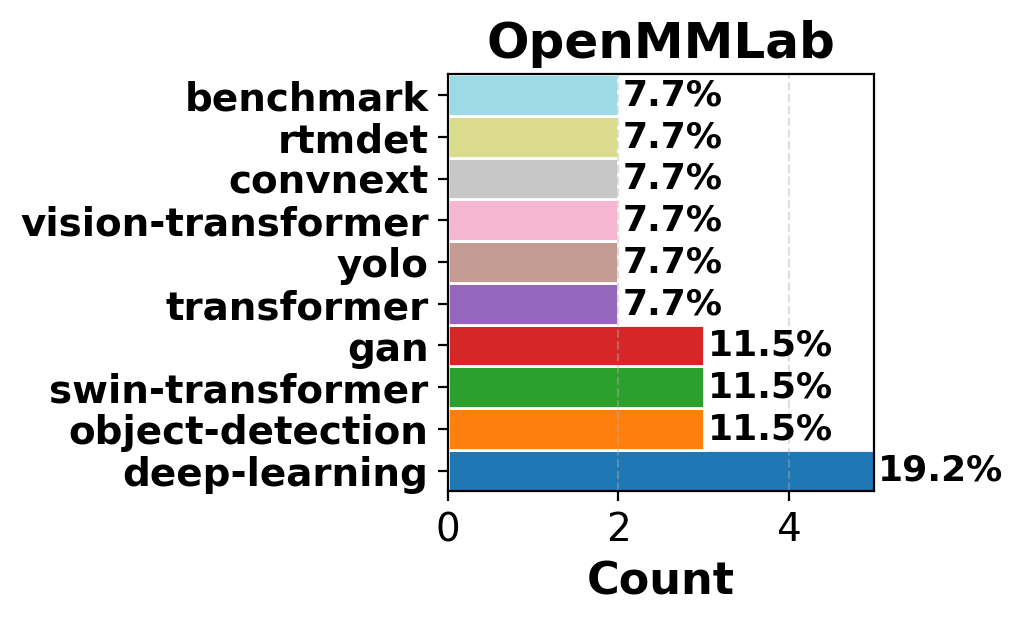}
    \end{minipage}\hfill
    \begin{minipage}[t]{0.19\textwidth}
        \includegraphics[width=\linewidth,keepaspectratio]{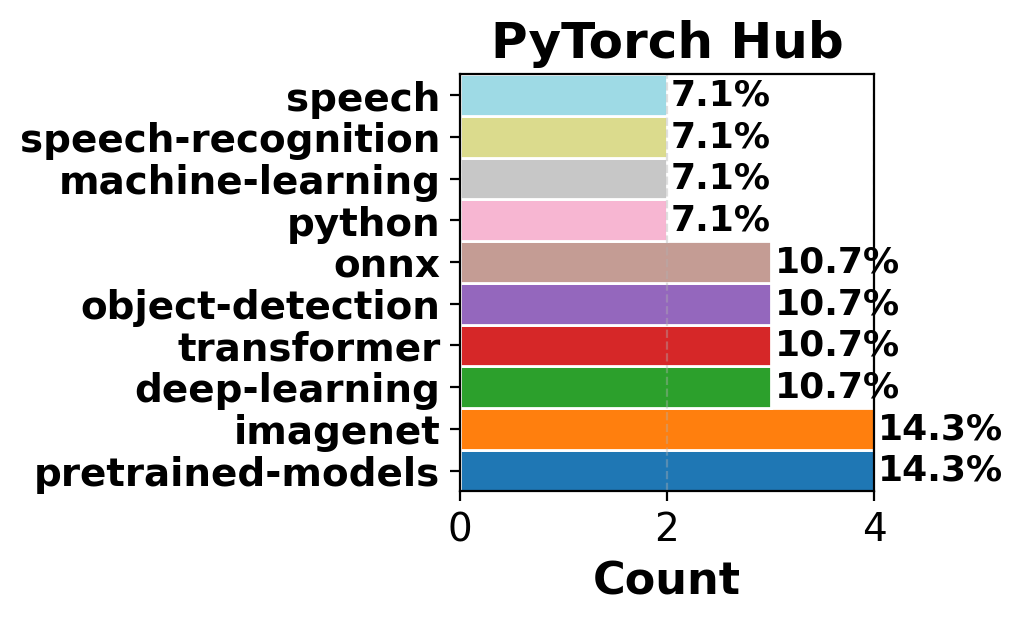}
    \end{minipage}

    \caption{Tag distributions of the custom models across different platforms.}
    \label{fig:combinedrq1result}
\end{figure}

Figure~\ref{fig:combinedrq1result} depicts the top 10 tag distributions of custom models across Hugging Face, ModelScope, OpenCSG, OpenMMLab, and PyTorch Hub platforms. For Hugging Face, the majority of custom models are used for \textit{text generation} (64.4\%), followed by \textit{feature extraction} (13\%). In contrast, for OpenCSG, around 21.8\% of the custom models are used for text generation, while roughly 15.1\% are associated with \textit{speech-related} tasks. Finally, for ModelScope, the distribution indicates that most custom models are concentrated in \textit{scientific and domain-specific applications}, reflecting its distinct usage patterns compared to the other two platforms. For PyTorch Hub, most models focus on vision, speech, and audio tasks, whereas in OpenMM Lab, model repositories mainly contain custom models for object detection and other vision tasks.

\subsection{RQ2: Security Analysis}
\subsubsection{Bandit Results}
Table~\ref{tab:rq2_bandit} summarizes the distribution of vulnerability types (CWE IDs) and the most frequent Bandit issues by severity across five model-sharing platforms. Across all platforms, \textbf{CWE-703 (Improper Check or Handling of Exceptional Conditions)} is the dominant weakness, accounting for over 75\% of findings on Hugging Face, OpenCSG, ModelScope, and OpenMMLab. This prevalence is largely driven by the low-severity Bandit rule \textbf{B101 (assert statements)}, which alone represents approximately 60--80\% of all reported issues. While typically low risk in isolation, these findings reflect fragile exception-handling practices that can undermine robustness and error containment.

On \textbf{Hugging Face}, 2,824 out of 36,697 repositories with custom code (7.70\%) exhibited at least one security smell. Beyond CWE-703, common weaknesses include \textbf{CWE-259 (Hard-coded Password)} and \textbf{CWE-494 (Download of Code Without Integrity Check)}. The most frequent high-severity issue was \textbf{B701} (Jinja2 autoescape disabled; XSS risk), while unsafe model downloads (\textbf{B615}) dominated medium-severity findings.

For \textbf{OpenCSG}, 31.92\% of repositories contained at least one detected issue. Similar to Hugging Face, CWE-703 and CWE-259 were prevalent; however, high-severity findings more frequently involved shell execution (\textbf{B602}), indicating elevated command-injection risk. Unsafe downloads (\textbf{B615}) again constituted the most common medium-severity issue.

\textbf{ModelScope} exhibits the lowest proportion of affected repositories (6.73\%), yet shows a nearly identical CWE profile dominated by CWE-703 and CWE-259. Low-severity B101 findings remain widespread, suggesting consistent shortcomings in defensive error handling.

Although \textbf{OpenMMLab} contains only 16 repositories, 75\% of them exhibit security smells. Here, CWE-703 accounts for 86.44\% of findings, while unsafe deserialization (\textbf{B301}) emerges as the most common high-severity issue, highlighting risks associated with loading untrusted artifacts.

Finally, \textbf{PyTorch Hub} shows a distinct vulnerability profile: \textbf{CWE-78 (OS Command Injection)} is the most frequent weakness, reflecting extensive use of shell commands and dynamic execution mechanisms. Nearly half of the repositories (46.15\%) contain security issues, confirming that this platform is particularly exposed to remote code execution risks.

\begin{table}[!htbp]
\centering
\scriptsize
\begingroup
\setlength{\aboverulesep}{0pt}
\setlength{\belowrulesep}{0pt}
\setlength{\tabcolsep}{2pt}
\renewcommand\cellalign{tl}
\renewcommand\cellgape{\Gape[1pt]}
\caption{Top 3 CWE and top 1 Bandit issue per severity across platforms.}
\label{tab:rq2_bandit}
\begin{tabular}{l c c l l}
\toprule
\textbf{Platform} & \textbf{\# Repos Analyzed} & \textbf{\# Smelly Repo (\%)} &
\textbf{Top 3 CWEs} & \textbf{Top 1 Issue per Severity (H, M, L)} \\
\midrule
Hugging Face & 36,697 & 2,824 (7.70\%) &
\makecell[l]{CWE-703 (Exception Handling): 76,535 (80.52\%)\\
CWE-259 (Hard-coded Password): 10,097 (10.62\%)\\
CWE-494 (No Integrity Check): 4,415 (4.64\%)} &
\makecell[l]{\textbf{H1}. B701 -- Jinja2 autoescape disabled (XSS risk). (56; 0.06\%)\\
\textbf{M1}. B615 -- Unsafe Hugging Face Hub download. (3,803; 4.00\%)\\
\textbf{L1}. B101 -- Use of assert detected. (74,298; 78.16\%)} \\
\midrule
OpenCSG & 6,165 & 1,968 (31.92\%) &
\makecell[l]{CWE-703 (Exception Handling): 14,393 (75.57\%)\\
CWE-259 (Hard-coded Password): 1,562 (8.20\%)\\
CWE-494 (No Integrity Check): 1,040 (5.46\%)} &
\makecell[l]{\textbf{H1}. B602 -- Starting a process with \texttt{shell=True}. (35; 0.18\%)\\
\textbf{M1}. B615 -- Unsafe Hugging Face Hub download. (983; 5.16\%)\\
\textbf{L1}. B101 -- Use of assert detected. (13,983; 73.42\%)} \\
\midrule
ModelScope & 3,193 & 215 (6.73\%) &
\makecell[l]{CWE-703 (Exception Handling): 17,293 (83.40\%)\\
CWE-259 (Hard-coded Password): 2,278 (10.99\%)\\
CWE-494 (No Integrity Check): 502 (2.42\%)} &
\makecell[l]{\textbf{H1}. B605 -- Starting a process with a shell. (12; 0.06\%)\\
\textbf{M1}. B615 -- Unsafe Hugging Face Hub download. (466; 2.25\%)\\
\textbf{L1}. B101 -- Use of assert detected. (17,033; 82.14\%)} \\
\midrule
OpenMMLab & 16 & 12 (75.00\%) &
\makecell[l]{CWE-703 (Exception Handling): 848 (86.44\%)\\
CWE-259 (Hard-coded Password): 124 (12.64\%)\\
CWE-502 (Unsafe Deserialization): 6 (0.61\%)} &
\makecell[l]{\textbf{H1}. B301 -- \texttt{pickle} deserialization of untrusted data. (4; 0.41\%)\\
\textbf{M1}. -- None observed.\\
\textbf{L1}. B101 -- Use of assert detected. (848; 86.44\%)} \\
\midrule
PyTorch Hub & 26 & 12 (46.15\%) &
\makecell[l]{CWE-78 (OS Command Injection): 42 (35.90\%)\\
CWE-703 (Exception Handling): 40 (34.19\%)\\
CWE-502 (Unsafe Deserialization): 20 (17.09\%)} &
\makecell[l]{\textbf{H1}. B605 -- Starting a process with a shell. (2; 7.69\%)\\
\textbf{M1}. B301 -- \texttt{pickle} deserialization of untrusted data. (1; 3.85\%)\\
\textbf{L1}. B101 -- Use of assert detected. (34; 30.77\%)} \\
\bottomrule
\end{tabular}
\endgroup
\end{table}

\subsubsection{Semgrep Results}
Table~\ref{tab:rq2_semgrep} presents the distribution of vulnerability types (CWE IDs), OWASP Top-10 categories, and the most frequent Semgrep rule violations across major model-sharing platforms. In contrast to Bandit which primarily surfaced low-severity coding smells, the Semgrep analysis reveals a strong concentration of \emph{security-critical issues}, particularly unsafe deserialization, dynamic code execution, and integrity violations.

Across all platforms, \textbf{CWE-502 (Deserialization of Untrusted Data)} is the most prevalent weakness, accounting for roughly \textbf{50--75\%} of all findings. This is consistently reflected in the dominance of Semgrep rules targeting unsafe \texttt{pickle} usage and PyTorch serialization patterns. Additional recurring weaknesses include \textbf{CWE-95 (Eval Injection)}, \textbf{CWE-706 (Incorrectly-Resolved Name or Reference)}, and \textbf{CWE-676 (Use of Potentially Dangerous Function)}, indicating widespread reliance on dynamic execution and unsafe imports.

The OWASP categories derived from Semgrep results align closely with these CWE patterns. \textbf{Injection} vulnerabilities dominate across platforms, accounting for 50.0\% of findings on Hugging Face and 67.9\% on PyTorch Hub, followed by \textbf{Insecure Deserialization} and \textbf{Integrity Failures}. This alignment highlights that many detected issues correspond to exploitable attack vectors rather than stylistic or defensive programming concerns.
\begin{table}[!htbp]
\centering
\scriptsize
\begingroup
\setlength{\aboverulesep}{0pt}
\setlength{\belowrulesep}{0pt}
\setlength{\tabcolsep}{2pt}
\renewcommand{\arraystretch}{1.05}
\setlength{\extrarowheight}{0pt}
\caption{Top 3 CWE, OWASP, and Semgrep issues across platforms.}
\label{tab:rq2_semgrep}

\begin{tabular}{l c l l l}
\toprule
\textbf{Platform} & \textbf{\# Smelly Repos (\%)} &
\textbf{Top 3 CWEs} & \textbf{Top 3 OWASP} & \textbf{Top 3 Rules} \\
\midrule

\multirow{3}{*}{Hugging Face} &
\multirow{3}{*}{1,691 (4.60\%)} &
CWE-502 (Unsafe Deserialization): 1,737 (65.45\%) &
Injection: 657 (50.04\%) &
numpy in pytorch: 1,946 (42.29\%) \\
& & CWE-95 (Eval Injection): 517 (19.48\%) &
Integrity Failures: 173 (13.18\%) &
pickles in pytorch: 1,557 (33.83\%) \\
& & CWE-79 (Cross-site Scripting): 96 (3.62\%) &
Insecure Deserialization: 171 (13.02\%) &
eval detected: 494 (10.73\%) \\
\midrule

\multirow{3}{*}{OpenCSG} &
\multirow{3}{*}{280 (4.54\%)} &
CWE-502 (Unsafe Deserialization): 502 (58.24\%) &
Injection: 169 (24.93\%) &
numpy in pytorch: 347 (28.65\%) \\
& & CWE-706 (Incorrectly-Resolved Name): 114 (13.23\%) &
Insecure Deserialization: 158 (23.30\%) &
pickles in pytorch: 328 (27.09\%) \\
& & CWE-95 (Eval Injection): 110 (12.76\%) &
Integrity Failures: 158 (23.30\%) &
avoid-pickle: 131 (10.82\%) \\
\midrule

\multirow{3}{*}{ModelScope} &
\multirow{3}{*}{509 (15.94\%)} &
CWE-502 (Unsafe Deserialization): 301 (70.16\%) &
Injection: 98 (60.87\%) &
numpy in pytorch: 327 (43.25\%) \\
& & CWE-95 (Eval Injection): 86 (20.05\%) &
Broken Access Control: 23 (14.29\%) &
pickles in pytorch: 283 (37.43\%) \\
& & CWE-706 (Incorrectly-Resolved Name): 23 (5.36\%) &
Insecure Deserialization: 18 (11.18\%) &
eval detected: 86 (11.38\%) \\
\midrule

\multirow{3}{*}{OpenMMLab} &
\multirow{3}{*}{2 (12.50\%)} &
CWE-502 (Unsafe Deserialization): 8 (61.54\%) &
Insecure Deserialization: 8 (38.10\%) &
avoid pickle: 8 (61.54\%) \\
& & CWE-95 (Eval Injection): 3 (23.08\%) &
Integrity Failures: 8 (38.10\%) &
eval detected: 2 (15.38\%) \\
& & CWE-706 (Incorrectly-Resolved Name): 2 (15.38\%) &
Injection: 3 (14.29\%) &
non-literal import: 2 (15.38\%) \\
\midrule

\multirow{3}{*}{PyTorch Hub} &
\multirow{3}{*}{10 (38.46\%)} &
CWE-502 (Unsafe Deserialization): 25 (49.02\%) &
Injection (A03:21): 19 (67.86\%) &
pickles in pytorch: 23 (46.94\%) \\
& & CWE-95 (Eval Injection): 16 (31.37\%) &
Injection (A01:17): 3 (10.71\%) &
eval detected: 16 (32.65\%) \\
& & CWE-676 (Dangerous Function): 5 (9.80\%) &
Insecure Deserialization: 2 (7.14\%) &
automatic memory pinning: 5 (10.20\%) \\
\bottomrule
\end{tabular}

\endgroup
\end{table}

On \textbf{Hugging Face}, 1,691 repositories (4.60\%) contain at least one Semgrep-detected issue, primarily driven by CWE-502 and CWE-95. \textbf{OpenCSG} exhibits a lower overall prevalence (4.54\%), yet similarly shows a CWE profile dominated by unsafe deserialization and dynamic execution. \textbf{ModelScope} shows a higher proportion of affected repositories (15.94\%), with CWE-502 accounting for nearly three-quarters of all findings. Although \textbf{OpenMMLab} contains only 16 repositories, 12.5\% exhibit Semgrep-detected security issues, largely related to unsafe deserialization. Finally, \textbf{PyTorch Hub} shows the highest relative impact (38.46\%), with CWE-502 and CWE-95 jointly accounting for over 80\% of detected weaknesses, underscoring a heightened exposure to remote code execution risks.

\subsubsection{CodeQL Results}
Table~\ref{tab:rq2_codeql} summarizes the CWE distribution and the most frequent CodeQL rule violations identified on Hugging Face, OpenCSG, and ModelScope. CodeQL did not report any findings for OpenMMLab or PyTorch Hub. Overall, CodeQL surfaces fewer affected repositories than Bandit and Semgrep, but highlights issues more closely tied to input handling, web security, and cryptographic practices.

Across platforms, \textbf{CWE-20 (Improper Input Validation)} and \textbf{CWE-79 (Cross-site Scripting)} appear consistently, while platform-specific weaknesses such as \textbf{CWE-22 (Path Traversal)} on Hugging Face and \textbf{CWE-730 (Regular Expression Denial of Service)} on OpenCSG are also observed. ModelScope is dominated by CWE-20, accounting for over three-quarters of its reported weaknesses, suggesting insufficient validation of external inputs.

The vast majority of CodeQL alerts across all platforms correspond to the \textbf{``Timing attack against secret''} query, which represents more than 95\% of all detections. This pattern indicates widespread use of non-constant-time comparisons in security-sensitive contexts, such as cryptographic checks or secret comparisons, potentially exposing repositories to side-channel attacks. Other cryptography-related queries, including generic cryptographic algorithm and hash usage checks, occur far less frequently.

\begin{table}[!htbp]
\centering
\scriptsize
\begingroup
\setlength{\aboverulesep}{0pt}
\setlength{\belowrulesep}{0pt}
\setlength{\tabcolsep}{2pt}
\renewcommand{\arraystretch}{1.05}
\setlength{\extrarowheight}{0pt}
\caption{Top 3 CWE and CodeQL issues across platforms.}
\label{tab:rq2_codeql}

\begin{tabular}{l c l l}
\toprule
\textbf{Platform} & \textbf{\# Smelly Repo (\%)} & \textbf{Top 3 CWEs} & \textbf{Top 3 Rules} \\
\midrule

\multirow{3}{*}{Hugging Face} & \multirow{3}{*}{1,894 (5.16\%)} &
CWE-79 (XSS): 59 (28.10\%) &
Timing attack against secret: 34,132 (98.48\%) \\
& & CWE-20 (Input Validation): 58 (27.62\%) &
All Cryptographic Algorithms: 99 (0.29\%) \\
& & CWE-22 (Path Traversal): 26 (12.38\%) &
Hash Algorithms: 99 (0.29\%) \\
\midrule

\multirow{3}{*}{OpenCSG} & \multirow{3}{*}{31 (0.50\%)} &
CWE-730 (ReDoS): 15 (26.79\%) &
Timing attack against secret: 4,705 (95.59\%) \\
& & CWE-20 (Input Validation): 10 (17.86\%) &
All Cryptographic Algorithms: 48 (0.98\%) \\
& & CWE-79 (XSS): 9 (16.07\%) &
Hash Algorithms: 48 (0.98\%) \\
\midrule

\multirow{3}{*}{ModelScope} & \multirow{3}{*}{161 (5.04\%)} &
CWE-20 (Input Validation): 18 (85.71\%) &
Timing attack against secret: 5,759 (99.12\%) \\
& & CWE-116 (Output Encoding): 3 (13.04\%) &
Overly permissive regular expression range: 18 (0.31\%) \\
& &  &
Weak hashes: 10 (0.17\%) \\
\bottomrule
\end{tabular}

\endgroup
\end{table}

On \textbf{Hugging Face}, 1,894 repositories (5.16\%) were flagged by CodeQL, with CWE-79, CWE-20, and CWE-22 accounting for the majority of identified weaknesses. \textbf{OpenCSG} shows a much lower prevalence (0.50\%), where ReDoS-related issues (CWE-730) are most prominent. \textbf{ModelScope} exhibits a similar overall impact (5.04\%), but with a CWE profile overwhelmingly dominated by improper input validation. Across all platforms, the dominance of timing-attack-related findings underscores a recurring risk stemming from unsafe cryptographic comparison patterns.

\subsubsection{YARA Results}
Table~\ref{tab:rq2_yara} summarizes YARA signature matches observed across model-sharing platforms. Overall, YARA findings affect a relatively small fraction of repositories; however, the detected signatures are dominated by \emph{environment and artifact identification rules} rather than explicit malware payloads. In particular, signatures associated with virtualized environments (\eg~Qemu, VBox, and VMWare detection) and specific file-format markers are the most prevalent across platforms.
\begin{table}[!htbp]
\centering
\scriptsize
\begingroup
\setlength{\aboverulesep}{0pt}
\setlength{\belowrulesep}{0pt}
\setlength{\tabcolsep}{2pt}
\renewcommand{\arraystretch}{1.05}
\setlength{\extrarowheight}{0pt}
\renewcommand\cellalign{tl}
\renewcommand\cellgape{\Gape[1pt]}
\caption{Top 3 YARA issues across platforms.}
\label{tab:rq2_yara}
\begin{tabular}{l c l}
\toprule
\textbf{Platform} & \textbf{\# Smelly Repo (\%)} & \textbf{Top 3 Rules} \\
\midrule
Hugging Face & 2,562 (6.98\%) &
JT 3D Visualization format: 24,778 (82.54\%)\\
& & VBox Detection: 1,260 (4.20\%)\\
& & Qemu Detection: 1,260 (4.20\%) \\
\midrule
OpenCSG & 24 (0.39\%) &
JT 3D Visualization format: 4,554 (60.55\%)\\
& & Suspicious signature: 648 (8.62\%)\\
& & VBox Detection: 492 (6.54\%) \\
\midrule
ModelScope & 200 (6.26\%) &
JT 3D Visualization format: 4,517 (86.28\%)\\
& & Qemu Detection: 164 (3.13\%)\\
& & VBox Detection: 164 (3.13\%) \\
\midrule
OpenMMLab & 12 (75.00\%) &
Suspicious signature: 38 (35.51\%)\\
& & TTA lossless compressed audio: 36 (33.64\%)\\
& & Audio Interchange File Format: 18 (16.82\%) \\
\midrule
PyTorch Hub & 3 (11.53\%) &
Qemu Detection: 6 (23.08\%)\\
& & VBox Detection: 6 (23.08\%)\\
& & VMWare Detection: 6 (23.08\%) \\
\bottomrule
\end{tabular}
\endgroup
\end{table}

On \textbf{Hugging Face}, 2,562 repositories (6.98\%) triggered at least one YARA rule. The vast majority of matches correspond to the \textbf{JT 3D Visualization format} signature (82.54\%), followed by virtualization-environment indicators such as \textbf{VBox Detection} and \textbf{Qemu Detection}. These signatures primarily indicate the presence of specific binary artifacts or environment checks rather than confirmed malicious behavior.

\textbf{OpenCSG} shows a very low prevalence of YARA hits (0.39\%), with detections again dominated by the \textbf{JT 3D Visualization format} rule (60.55\%), alongside generic \textbf{Suspicious signature} and \textbf{VBox Detection} rules. \textbf{ModelScope} exhibits a similar pattern, with 6.26\% of repositories flagged and over three-quarters of matches attributed to the JT file-format signature, followed by Qemu and VBox detections.

Although \textbf{OpenMMLab} contains only 16 repositories, 75\% of them triggered YARA rules. Unlike other platforms, its top signatures correspond to generic \textbf{Suspicious signature} indicators and audio file–format signatures (TTA and AIFF), reflecting the presence of uncommon binary artifacts rather than virtualization checks.

Finally, \textbf{PyTorch Hub} shows YARA matches in 11.53\% of repositories, with detections evenly distributed across \textbf{Qemu}, \textbf{VBox}, and \textbf{VMWare Detection} rules. Across all platforms, the dominance of environment-detection and file-format signatures suggests that YARA primarily surfaces \emph{artifact-level indicators} rather than clear evidence of embedded malware.

\subsection{RQ3: Platform Mitigation Strategies}
\label{subsec:rq3_results}

Table~\ref{tab:platform_security_mechanisms} summarizes the security mechanisms of each platform. 

\begin{table}[!htbp]
\centering
\scriptsize
\setlength{\aboverulesep}{0pt}
\setlength{\belowrulesep}{0pt}
\setlength{\tabcolsep}{2pt} 
\renewcommand{\arraystretch}{1.05} 
\setlength{\extrarowheight}{0pt}   
\renewcommand\cellalign{tl}
\renewcommand\cellgape{\Gape[1pt]} 
\caption{Comparison of security and trust mechanisms across model-sharing platforms verified against official documentation.}
\label{tab:platform_security_mechanisms}
\begin{tabularx}{\textwidth}{l X X X X X}
\toprule
\textbf{Platform} & \textbf{Upload Verification} & \textbf{Trust Model} & \textbf{Malware Scanning} & \textbf{Warning Systems} & \textbf{User Protection} \\
\midrule
Hugging Face &
Automated scanning~\cite{hf2024security,hftrufflesecurity2024}
&
Trust-all with verified badges~\cite{hforgverification2024} &
Yes -- comprehensive multi-layered~\cite{hfmalwarescan2024,ciscoclamav2024} &
UI warnings, file badges (ok/infected), \& verified badges~\cite{hfmalwarescan2024} &
Documentation hub~\cite{hfsecurity2024}, UI warnings, SafeTensors support~\cite{hf2024security} \\
\midrule
OpenCSG &
Open uploads via Git or web~\cite{opencsgupload2024}
&
Community trust~\cite{csghubcommunity2024} &
None documented~\cite{opencsgdocs2024} &
None documented~\cite{opencsgdocs2024} &
Documentation (community guidelines)~\cite{opencsgdocs2024} \\
\midrule
ModelScope &
No platform-level automated scanning documented~\cite{modelscoperepo2024} &
Trust-all (no verification)~\cite{modelscopephi3} &
None -- no platform-level scanning~\cite{modelscoperepo2024} &
Post-download \texttt{trust\_remote\_code} warnings only~\cite{evalscopefaq2024,modelscoperelease2024} &
Documentation, \texttt{trust\_remote\_code} parameter~\cite{modelscoperelease2024} \\
\midrule
OpenMMLab &
Maintainer pull-request review~\cite{mmdet3dcontrib2024} &
Verify-first~\cite{mmdet3dcontrib2024} &
None documented~\cite{mmdetmodelzoo2024} &
None documented~\cite{mmdetmodelzoo2024} &
Documentation (reviewed code)~\cite{mmdetrepo2024} \\
\midrule
PyTorch Hub &
No automated verification~\cite{pytorchhub2024} 
&
Trust-all with \texttt{trust\_repo} parameter~\cite{pytorchhub2024} &
None documented~\cite{pytorchhub2024} &
Interactive prompts (\texttt{trust\_repo}), deprecation warnings~\cite{pytorchhub2024,pytorchissue52181} &
Strong documentation warnings: ``models are programs''~\cite{pytorchsecurity2024} \\
\bottomrule
\end{tabularx}
\end{table}

\subsubsection{Trust Models and Verification}
Platforms exhibit three distinct trust paradigms. \textbf{Hugging Face}, \textbf{ModelScope}, and \textbf{PyTorch Hub} follow \emph{trust-all} models where any user can freely upload models. Hugging Face augments this with verified badges for organizational identity (not security audits)~\cite{hforgverification2024}, while PyTorch Hub shifts trust decisions to users via the \code{trust\_repo} parameter~\cite{pytorch2025}. \textbf{OpenMMLab} implements strict \emph{verify-first} with maintainer review of all contributions through pull requests~\cite{mmdet3dcontrib2024}. \textbf{OpenCSG} represents a middle ground with \emph{community trust}, allowing open uploads with optional content moderation~\cite{opencsgupload2024,csghubserver2024}.

Only \textbf{Hugging Face} operates comprehensive automated security scanning, triggering on every push with ClamAV (malware), PickleScan (pickle/RCE), TruffleHog (secrets), plus third-party scanners (Protect AI Guardian, JFrog)~\cite{hf2024security,hftrufflesecurity2024,hfazuresecurity2024}. However, scans target known patterns rather than comprehensive static analysis, leaving residual RCE risks~\cite{hfmalwarescan2024,hfsecretsscanning2025}. \textbf{OpenCSG}~\cite{opencsg2025}, \textbf{ModelScope}~\cite{modelscope2025}, and \textbf{PyTorch Hub}~\cite{pytorch2025} have no documented platform-level automated scanning. \textbf{OpenMMLab} relies on human review without automated scanning~\cite{mmdetmodelzoo2024}. No platform implements comprehensive sandboxing for custom code execution during model loading.

\subsubsection{User-Facing Protections}
\textbf{Hugging Face} provides the most comprehensive user protections: verified organizational badges~\cite{hfmalwarescan2024}, prominent UI banners for \code{trust\_remote\_code=True}, file badges (ok/infected), a dedicated security documentation hub~\cite{hfsecurity2024}, and SafeTensors support~\cite{hf2024security}. However, model card code snippets lack inline warnings about \code{trust\_remote\_code} risks. \textbf{PyTorch Hub} has documentation warnings emphasizing that ``models are programs''~\cite{pytorchsecurity2024} with interactive prompts via \code{trust\_repo}~\cite{pytorch2025,pytorchissue52181}. \textbf{ModelScope} provides only post-download \code{trust\_remote\_code} warnings~\cite{evalscopefaq2024,modelscoperelease2024}. \textbf{OpenCSG} includes community guidelines~\cite{opencsgdocs2024} but lacks explicit \code{trust\_remote\_code} warning documentation. \textbf{OpenMMLab} has no warnings due to its curated, reviewed model zoo~\cite{mmdet3dcontrib2024,mmdetrepo2024}.

\subsection{RQ4: Developers' Concern}
\begin{figure}[!htbp]
    \centering
    \includegraphics[width=\linewidth]{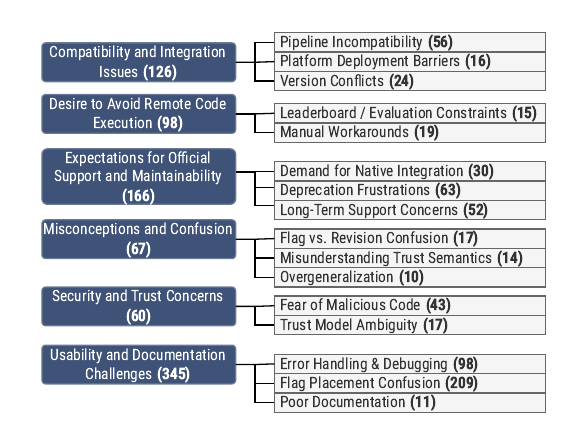}
    \caption{Taxonomy of Developers' Concerns.}
    \label{fig:rq4_taxonomy}
\end{figure}

From our systematic analysis of developers' discussions surrounding the \texttt{trust\_remote\_code} mechanism, we derived a taxonomy of concerns observed across forums, issue trackers, and community posts (Figure \ref{fig:rq4_taxonomy}). It is worth noting that while each main concept is decomposed into finer-grained sub-concepts, we also observed posts that express a high-level concern but do not clearly fit into any sub-concept. These posts are therefore assigned directly to the main concept, meaning that aggregate counts at the main-concept level may exceed the sum of their sub-concept counts. 

\paragraph{Compatibility and Integration Issues.}
This category represents discussions where developers report integration failures, incompatibilities, or unexpected behavior when enabling the trust flag. These posts often feature vague complaints such as \textit{``it doesn’t work''}, \textit{``the model fails to load''}, or \textit{``no support for this flag''}, typically accompanied by minimal debugging information. 

\begin{itemize}
    \item \textbf{Pipeline Incompatibility:} A recurring sub-theme involves failures when loading models through pipelines or inference APIs. Because the trust flag is not consistently propagated through these abstractions, developers experience silent failures or partial functionality, requiring non-trivial debugging. These cases often reveal architectural gaps between the core model loaders and downstream pipeline wrappers.
    \item \textbf{Platform Deployment Barriers:} Many developers face difficulties deploying trust-dependent models on managed hosting platforms such as SageMaker, Inference Endpoints, or custom cloud containers. Restrictions on executing remote code, security sandboxing, and the lack of explicit trust flag support in deployment configurations contribute to deployment dead ends.
    \item \textbf{Version Conflicts:} Another dominant sub-pattern involves versioning issues. Breakages often occur due to mismatches between the installed library version and the model’s expected environment. Posts frequently cite outdated \texttt{transformers} packages, missing backward compatibility, or changes in the trust flag’s default behavior across versions. These issues can cascade as small version drifts can break complex pipelines.
\end{itemize}

\paragraph{Desire to Avoid Remote Code Execution.}
This category captures discussions where developers explicitly express their reluctance or refusal to enable the trust flag. Unlike compatibility issues, these concerns stem from security or policy perspectives, or from a general mistrust of executing third-party code. 

\begin{itemize}
    \item \textbf{Leaderboard / Evaluation Constraints:} In competitive or benchmarking contexts, enabling trust flags is sometimes explicitly forbidden. This restriction stems from fairness, reproducibility, or sandboxing requirements, forcing developers to look for alternative workarounds.
    \item \textbf{Manual Workarounds:} Developers frequently fork repositories, manually download and edit model files, or patch library internals to bypass trust requirements. While these ad-hoc solutions may allow immediate progress, they introduce technical debt, security uncertainty, and maintenance challenges downstream.
\end{itemize}

\paragraph{Expectations for Official Support and Maintainability.}
We found discussions that showed expectations from the community for upstream maintainers and platform providers to \textit{``just make it work''}. This category reflects the \textbf{expectation gap} between what developers assume the trust flag should offer (automatic, safe, supported execution) and what is actually implemented (manual flag toggling, fragmented support, and inconsistent documentation).

\begin{itemize}
    \item \textbf{Demand for Native Integration:} Developers requested that maintainers integrate model-specific custom code directly into official libraries, thereby removing the need for explicit trust flags. This reflects a preference for official, standardized mechanisms over user-managed trust settings.
    \item \textbf{Deprecation Frustrations:} As the trust mechanism and related APIs evolve, developers face broken pipelines and inconsistent behavior. Complaints in this sub-category often highlight insufficient deprecation notices, breaking changes without migration guides, and a lack of backward compatibility.
    \item \textbf{Long-Term Support Concerns:} Developers working in production environments or regulated domains express concern over whether trust-based model integrations will remain viable in the future. These concerns are often tied to compliance, maintenance, and stability over multiple product cycles.
\end{itemize}

\paragraph{Misconceptions and Confusion.}
Not all developer challenges arise from genuine technical limitations. Some stemmed from an incomplete or incorrect understanding of how \texttt{trust\_remote\_code} operates. 

\begin{itemize}
    \item \textbf{Flag vs. Revision Confusion:} Developers often conflate the trust flag with revision pinning or version control, mistakenly believing that setting a revision automatically enables trust or vice versa.
    \item \textbf{Misunderstanding Trust Semantics:} Many users incorrectly assume that enabling the flag merely grants permission for metadata loading, not remote code execution. This misinterpretation may lead to underestimating security implications or failing to configure environments correctly.
    \item \textbf{Overgeneralization:} Another common misconception involves assuming that the trust flag behaves uniformly across all model architectures and frameworks. In practice, its support is uneven, leading to mismatched expectations and implementation failures.
\end{itemize}

\paragraph{Security and Trust Concerns.}
It represents a distinct and high-stakes theme in developer discourse. Here, developers explicitly reference potential or perceived security risks associated with enabling trust flags. Unlike the “Desire to Avoid RCE” category, which is attitudinal, this category focuses on \textbf{explicit threat articulation}.

\begin{itemize}
    \item \textbf{Fear of Malicious Code:} Developers express concerns about arbitrary code execution, supply chain compromises, or untrusted contributors injecting malicious payloads. These discussions frequently reference standard security practices, organization-level security policies, or compliance concerns.
    \item \textbf{Trust Model Ambiguity:} Many developers do not fully understand what “trusting” a model entails at the technical level (\eg~which parts of the repository are executed, what isolation exists, or what verification is done). This lack of transparency fuels suspicion and defensive behavior.
\end{itemize}

\paragraph{Usability and Documentation Challenges.}
Even when the trust mechanism works as intended, poor documentation, unclear error messages, or confusing flag placement can create technical barriers. 

\begin{itemize}
    \item \textbf{Error Handling \& Debugging:} Many developers encounter non-informative or misleading error messages when enabling or failing to enable the trust flag. These debugging hurdles often prolong troubleshooting cycles.
    \item \textbf{Flag Placement Confusion:} Developers frequently struggle to identify where the trust flag should be set (\eg in CLI arguments, in \texttt{pipeline} calls, or at model initialization), especially when documentation is inconsistent across versions.
    \item \textbf{Poor Documentation:} We found posts citing missing, incomplete, or outdated documentation, a lack of minimal working examples, and inconsistent terminology. 
\end{itemize}

\section{Discussion}\label{sec:discussion}

\subsection{Ecosystem-wide Security Exposure}
Our findings (Table~\ref{tbl:rq1_results}) show that the \textbf{model-sharing ecosystem is broadly and unevenly exposed to security risks}. While only 2--4\% of models on platforms such as Hugging Face, ModelScope, and OpenCSG require custom code, this seemingly small subset represents approximately \emph{45,000 repositories} containing code that executes at model load time. Platforms such as OpenMMLab and PyTorch Hub rely almost entirely on custom code, substantially increasing their systemic attack surface. This volume is further exacerbated by rapid hype cycles, where a single landmark release often triggers thousands of community-led derivatives, such as fine-tunes and quantizations, each potentially propagating or introducing insecure code-execution paths at scale.

Static analysis with Bandit and Semgrep reveals two dominant vulnerability clusters. First, \textbf{low-severity but pervasive coding smells} (e.g., CWE-703 and Bandit B101 \texttt{assert} statements) appear in 60--80\% of affected repositories, reflecting weak defensive programming and exception-handling practices. Second, \textbf{high-impact injection and deserialization vulnerabilities} (e.g., CWE-502, CWE-95, CWE-78) are widespread, particularly on Hugging Face and PyTorch Hub, where dynamic execution via \texttt{pickle}, \texttt{eval}, and shell commands is common. Semgrep analysis further highlights \textbf{Injection} and \textbf{Insecure Deserialization} as the dominant OWASP categories, underscoring systemic risks of arbitrary code execution during model loading.

Our CodeQL analysis reinforces these observations while providing complementary coverage of \emph{input handling and cryptographic practices} (Table~\ref{tab:rq2_codeql}). Across platforms, the vast majority of CodeQL alerts (over 95\%) correspond to the ``Timing attack against secret'' query, indicating widespread use of non-constant-time comparisons or insecure secret handling. Although these findings may not always translate to immediately exploitable vulnerabilities, their prevalence reflects \emph{weak default security hygiene} in model repository codebases. In addition, recurring weaknesses such as \textbf{CWE-20 (Improper Input Validation)} and \textbf{CWE-79 (Cross-site Scripting)} suggest that many repositories lack basic safeguards against malformed or adversarial inputs, particularly in web-facing or configuration-driven components.

Importantly, security exposure is not uniform across platforms. OpenCSG contributes a high absolute number of findings due to its scale, while \textbf{PyTorch Hub}, despite its smaller ecosystem, exhibits a disproportionately high rate of high-severity issues, especially command injection and unsafe deserialization. These differences highlight the role of platform trust boundaries, review practices, and default loading mechanisms in shaping real-world security risk.

\subsection{Gaps Between Security Mechanisms and Developer Practices}
The analysis of platform security mechanisms (Table~\ref{tab:platform_security_mechanisms}) reveals a clear \textbf{misalignment between available safeguards and developer practices}. Hugging Face, for example, operates the most advanced malware scanning and warning pipeline among the studied platforms, yet \emph{unsafe coding patterns persist widely}, including reliance on \texttt{pickle} serialization, dynamic execution, and unpinned revision loading. The high prevalence of CWE-502 and CWE-95 demonstrates that \textbf{technical defenses alone are insufficient to change developer behavior}.

Similarly, ModelScope issues warning banners for \texttt{trust\_remote\_code} but lacks sandboxing or pre-upload verification, allowing risky code to propagate unchecked. OpenCSG and PyTorch Hub provide only lightweight defenses, such as trust prompts or community-based assumptions, without systematic static or dynamic analysis. The concentration of injection- and eval-related vulnerabilities on PyTorch Hub illustrates the risks of such minimal safeguards.

Finally, the \emph{low adoption of safer alternatives}, such as secure serialization formats (\eg~Safetensors), suggests that usability, ecosystem inertia, and lack of incentives continue to outweigh security considerations. As a result, even when safer mechanisms exist, they are rarely adopted at scale, leaving the model-sharing ecosystem exposed to avoidable risks.



\subsection{Results Implications}
Our findings have important implications for platform operators, developers, and the research community.

\paragraph{For Platform Operators}
Platform operators should move beyond passive warning mechanisms toward \textbf{enforced security boundaries}. This includes default sandboxing of untrusted custom code, mandatory integrity and provenance checks, and stricter upload-time verification workflows. Providing richer developer-facing telemetry, such as inline vulnerability alerts, dependency provenance, and risk summaries at load time, can help bridge the gap between automated scanning and the practical adoption of secure development practices.

\paragraph{For Developers and Maintainers}
The results highlight that developers and maintainers play a decisive role in shaping the security posture of model repositories. Reliance on unsafe mechanisms such as \texttt{pickle}-based deserialization and \texttt{eval}-driven execution should be minimized or replaced with safer loading alternatives whenever possible. Adopting secure defaults, revision pinning, and lightweight security review checklists can substantially reduce the prevalence of recurring weaknesses, particularly CWE-502 and CWE-95.

\paragraph{For Researchers}
Our study shows that, despite relying on shared underlying libraries, model-sharing platforms differ substantially in how they handle custom code during model loading. This points to the need for research on \textbf{automated enforcement frameworks} that explicitly define and enforce trust boundaries, combining cryptographic integrity verification with runtime isolation. Moreover, our results indicate that tools such as CodeQL can \textbf{reveal deep structural weaknesses} in model repository ecosystems that are not surfaced by conventional scanners alone. This creates opportunities to design integrated approaches that couple vulnerability detection with upload-time checks, runtime sandboxing, and integrity enforcement. Future work should also investigate adoption barriers for secure coding practices, particularly around cryptographic operations, to close the gap between warnings and actionable defenses.


\subsection{Threats to Validity}


\paragraph{Internal Validity.} 
A primary internal threat lies in the accuracy and completeness of our static analysis. Although we employed three well-established tools—Bandit, CodeQL, and Semgrep—to identify security smells and CWE patterns, they may produce false positives or false negatives. However, Siddiq \etal show Bandit has 90.79\% precision \cite{siddiq2022empirical}. Semgrep, CodeQL, and YARA are widely used in the research community \cite{siddiq2022seceval,gnieciak2025largelanguagemodelsversus,jåtten2025scalablethreadsafetyanalysisjava,bennett2024semgrep,kree2024using,siddiq2024sallm, naik2020evaluating, naik2021embedded}. Moreover, we manually analyzed discussion posts to conduct open-coding. As mentioned before, this coding was conducted collaboratively by the two authors, whose software development experience ranged from 4 to 12 years, with disagreements resolved by the senior author. The Cohen's kappa score is 0.50, indicating a “moderate” level of agreement \cite{cohen1960kappa}. 
\paragraph{External Validity.} 
Our results may not fully generalize beyond the platforms studied. We focused on five major platforms—Hugging Face, OpenCSG, ModelScope, OpenMMLab, and PyTorch Hub—that dominate the model-sharing ecosystem. For example, Hugging Face hosts around 1.7 million models, and OpenCSG hosts around 200k models. Smaller or private repositories (e.g., enterprise model registries) may exhibit different security characteristics. 

\paragraph{Construct Validity.}
A threat to construct validity stems from the use of static reachability analysis to approximate execution during model loading. Although our JARVIS-based analysis restricts findings to code paths reachable from model-loading entry points, Python’s dynamic features (\eg~reflection, dynamic imports, and runtime code generation) may not be fully resolved statically. Consequently, our analysis may under-approximate the true execution surface in rare cases. We mitigate this threat by grounding entry-point selection in files associated with custom model code and by using a flow- and context-sensitive call-graph analysis; nevertheless, we conservatively interpret our results as lower-bound estimates of execution-relevant security exposure.
\section{Related Work}
\label{sec:related}

\subsection{Evolution of Model Hosting Platforms and Pipelines}
\label{sec:related:evolution}
The evolution from localized model development to centralized sharing platforms constitutes a shift to collaborative ML practices. In the early stages, researchers relied on manual distribution via institutional websites or GitHub repositories, requiring end users to rebuild the entire training and execution environment to reproduce results. The first generation of organized model distribution are mainly \textit{\href{https://caffe.berkeleyvision.org/}{Caffe}} (2014) and \textit{TensorFlow Hub} (2018). With \textit{PyTorch Hub} created in 2019, the \texttt{torch.hub.load()} interface was also released along with the \texttt{trust\_repo} parameter~\cite{pytorchhub2024}. The rapid expansion of \textit{Hugging Face} between 2018 and 2023 further reshaped the landscape: evolving from a simple hosting repository to a fully integrated platform supporting training, inference, and deployment workflows. Zhao \etal~\cite{Jian24modelsarecode} provides a comprehensive analysis of this ecosystem’s development, showing how model hubs have become critical infrastructure sustaining millions of models and billions of downloads worldwide. This infrastructural transformation is further quantified by Laufer \etal, by analyzing two million models hosted on Hugging Face~\cite{laufer2025anatomy}. Their findings highlight the platform’s support for over 4,000 distinct architectures, with an increasing proportion of models depending on custom code to enable advanced functionality beyond standard implementations. Our work focuses on the architectural design of executing code during model loading from the hub.

\subsection{Quality and security issues of Model Hubs}
Jiang \etal~\cite{jiang2022artifacts} conducted the first systematic study of these artifacts across eight platforms, revealing that trust relationships in model ecosystems are more implicit and poorly understood than in traditional software. Hu \etal.~\cite{hu2025llmsupplychain} identifies open problems in the LLM supply chain, documenting how fine-tuning workflows, adapter layers, and prompt templates all serve as injection points. Yi \etal~\cite{ma2025understandingsupplychainrisks} further characterizes these risks from an edge-computing perspective, showing how LLM-integrated platforms create new trust boundaries among cloud services, edge devices, and end users. 

The introduction of \code{weights\_only=True} in PyTorch 1.13~\cite{pytorchissue52181} and the \code{trust\_remote\_code} flag in Transformers 4.0~\cite{huggingface2025trust} represent acknowledgments of the risks, but adoption remains low due to compatibility concerns. Alternative serialization approaches exhibit different trade-offs between security and functionality. Safetensors, introduced by Hugging Face in 2022~\cite{safetensorsdocs2025}, uses a simple header-data format that prevents code execution entirely. The format stores tensors in a flat layout with minimal metadata, enabling zero-copy loading while eliminating executable payloads. However, as our results show and Laufer \etal.\ confirm~\cite{laufer2025anatomy}, only 6.6\% of models have adopted this format despite platform encouragement. Recent work on secure deserialization provides partial solutions. 

PickleBall~\cite{kellas2025pickleball} introduces semantics-aware loading that permits safe pickle subsets through allowlisting specific opcodes and validating operation sequences. Related vulnerabilities in other frameworks, such as CERT/CC's advisory on Keras Lambda layers (CVE-2024-3660)~\cite{cert2024keras}, demonstrate that the problem extends beyond pickle to any format that conflates data with executable logic.

While platform owners use scanners to identify vulnerable code and data, Zhao \etal's deployment of \emph{MalHug}~\cite{Jian24modelsarecode} identified 91 malicious models and 9 dataset scripts actively exploiting users. JFrog Security Research~\cite{jfrog2024silent} documented evasion methods that bypass pattern-based scanning, including time-delayed execution, environment fingerprinting, and polymorphic code generation. The August 2025 Protect AI report~\cite{protectai2025sixmonths}, based on scanning 4.47 million model versions, identifies emerging threats, such as archive slip vulnerabilities and TensorFlow-specific backdoors, that existing tools miss. Our work specifically focuses on the code associated with the model, which is executed during loading, and on developers' concerns about it.
        
\subsection{Mitigation Strategies and Their Limitations}
\label{sec:related:mitigation}
Current mitigation strategies operate at multiple levels with varying effectiveness. Platform-level approaches include automated scanning, trust indicators, and secure format promotion. Hugging Face's multi-layered architecture~\cite{hfmalwarescan2024,hfpicklescan2024,hfsecretsscanning2025} combines ClamAV for malware, custom pickle analysis, secrets scanning via TruffleHog~\cite{hftrufflesecurity2024}, and third-party integration with Protect AI Guardian~\cite{hfprotectai2024}. The Cisco partnership adds enterprise-grade threat detection~\cite{ciscoclamav2024}, processing 226 million scan requests monthly as of April 2025~\cite{protectai2025sixmonths}. Technical solutions at the framework level show promise but lack adoption. SafeTensors~\cite{safetensorsdocs2025,hfsafetensors2025} eliminates serialization-based attacks entirely, yet faces resistance due to compatibility concerns and performance assumptions (despite offering comparable or better loading speeds). However, these platform mitigation strategies face fundamental limitations. Static analysis struggles with Python's dynamic features and cannot reason about runtime behavior. Signature-based detection fails against novel threats and polymorphic malware. Most critically, scanning occurs after upload, meaning malicious models can execute before detection. Our analysis of platform documentation reveals that only Hugging Face provides comprehensive scanning, while ModelScope, OpenCSG, and PyTorch Hub offer only minimal automated verification.

\subsection{Adjacent Threats: Model-Level Attacks}
\label{sec:related:adjacent}
While our focus is loader-time host compromise, model-level attacks represent complementary threats in the same ecosystem. BadNets~\cite{gu2017badnets} pioneered neural backdoors that activate on specific triggers, while subsequent work~\cite{liu2018trojaning} demonstrated backdoors surviving transfer learning and model compression. Other works similarly investigate data poisoning attacks and how introducing trigger phrases or keywords can alter the output of a model to be something that an attacker specified \cite{wallace2021concealed, kurita2020weight, 9581257, 9402020}. These attacks differ from code execution in that they compromise model behavior rather than host systems, but they exploit the same distribution channels and trust relationships.

Recent surveys~\cite{li2025llmsurvey} catalog the expanding threat landscape for large language models: prompt injection, training data extraction, adversarial examples, and model inversion attacks. These behavioral compromises often combine with supply chain attacks- a Trojan model might also contain loader malware, maximizing adversarial impact. The convergence of multiple threat vectors reinforces our finding that model-sharing platforms must address security holistically rather than focusing on individual attack types.

\section{Conclusion and Future Work}\label{sec:conclusion}
Our work provides the first large-scale, cross-platform empirical analysis of remote code execution risks in ML model hosting ecosystems, examining five major platforms. We identified that around 45,000 repositories execute arbitrary code during model loading. Our static analysis revealed most repositories have weak defensive coding practices, and injection and deserialization vulnerabilities (\eg CWE-502, CWE-95, CWE-78. We also found  that most of the malicious code in the category of environmental evasion indicators (\eg~Qemu, VMWare, VBox detections). Although Hugging Face has made significant advances with automated malware scanning pipelines (\eg~ClamAV, PickleScan, Protect AI Guardian), these mechanisms alone are insufficient. Other platforms lack comparable safeguards, with minimal or no sandboxing and weak verification mechanisms. Developer discussions further reveal widespread confusion and misconceptions about trust flags, limited adoption of secure serialization formats like SafeTensors, and tension between usability and security. In the future, we will focus on an automated enforcement framework for trust boundaries, integrating cryptographic integrity verification with runtime isolation for remote code execution during model loading. We will also explore safe alternatives to custom model loading.

\bibliographystyle{ACM-Reference-Format}
\bibliography{references}


\end{document}